\documentstyle[aps,prb,multicol]{revtex}
\draft
\begin{document}
\bibliographystyle{unsrt}
\title{ Composite quasiparticle formation and
the low-energy
effective Hamiltonians of  the
one- and two-dimensional Hubbard Model
}
\author{ Junwu Gan and Dung-Hai Lee }

\address{ Department of Physics,
	University of California, Berkeley, CA 94720.}
\author{ Per Hedeg\aa{}rd}
\address{\O{}rsted Laboratory, Niels Bohr Institute,
 University of Copenhagen, \\
Universitetsparken 5, DK-2100 Copenhagen \O{}, Denmark}

\maketitle

\begin{abstract}

We investigate the effect of hole doping on the strong-coupling
Hubbard model at half-filling in spatial dimensions $D\ge 1$.
We start with an
antiferromagnetic mean-field description of the insulating state, and
show that doping creates solitons in the antiferromagnetic background.
In one dimension, the soliton is topological,
spinless, and decoupled from the
background
antiferromagnetic fluctuations at low energies.
In two dimensions and above, the
soliton is non-topological, has
spin quantum number 1/2, and is strongly coupled
to the antiferromagnetic
fluctuations.
We derive the effective action governing the quasiparticle motion, study
the properties of a single carrier, and comment on a possible
description
at finite concentration.

\end{abstract}

%\vskip 20pt

\pacs{PACS Numbers: 71.10.Fd, 74.20.Mn, 71.27.+a }

%======================================================
			% BODY OF PAPER
%\begin{multicols}{2}

\section{Introduction}

Nearly sixty years ago, Peierls first suggested that
correlation effects are the cause of the insulating behavior in
nickel oxide.\cite{mott} Today, we know that NiO is an example
of a ``Mott insulator'' -- an insulator that would be metallic
by valence counting.  The parent compounds of high temperature
superconductors, e.g., La$_2$CuO$_4$, are well
described as Mott insulators, and have stimulated a renewed
interest in the subject.

Mott insulators are generally antiferromagnetic, a feature which
makes a mean-field spin-density-wave (SDW)
description possible\cite{slater}.
Although an antiferromagnetic insulator within
the SDW description resembles a band insulator,
the difference
between the two types of insulators is most pronounced when charge
carriers are added.  Doping a band insulator (say with $p$-type
dopants) moves the Fermi energy from midgap into the valence band.
The hole ``pocket'' in the valence band is responsible for the
conductivity.  In a Mott insulator, however, doping creates solitons:
localized regions of suppressed antiferromagnetic order
in which the hole is confined. The
energies of the soliton states lie inside the SDW gap and
pin the chemical potential.
The soliton propagation is
responsible for the conductivity.

The nature of the solitons in Mott insulators depends on the spatial
dimensionality.  In one dimension (1D), the solitons are
topological -- they are the anti-phase domain wall.\cite{shiba,lee90}
They have spin quantum number zero and are decoupled
from the spin fluctuations at low energies.
As a result of the decoupling, the quasiparticle
bandwidth is only weakly
renormalized.\cite{lee90} In $D\ge 2$, the
solitons are non-topological -- they are ``bag''-like
objects in which the antiferromagnetic order is suppressed
consistent with the symmetry of the underlying
lattice.\cite{schrieffer88,su88,mele88} They carry spin and
interact strongly with the low-energy spin fluctuations,
significantly reducing
the quasiparticle bandwidth.

Both numerical evidence and the behavior of the optical
conductivity in the high temperature superconductors are
suggestive of the existence of such quasiparticles in
two dimensions.  Numerical
solutions for the  single particle Green's function indicate a
single pole and an incoherent
background.\cite{dago94,dago90,poil93,eder94,haas94,moreo94}
In the high $T_c$ oxides, the optical
conductivity \cite{optical} contains two
components, a Drude peak and a broad mid-infrared absorption,
which are related to the added charge
carriers (their spectral weight satisfies the sum rule)
and can not be well fit over the entire region
assuming conventional quasiparticles.
Furthermore, the two components have very different
temperature dependence.
While the
Drude peak is sensitive to
temperature (e.g. the width of the
Drude peak is $\approx 2k_BT$), the
mid-infrared part is not.
In fact, the latter survives even in the
superconducting state.
This type of optical conductivity is, however, what
would be expected from
solitonic quasiparticles whose conduction
is responsible for the Drude peak and whose internal structure
generates the mid-infrared absorption.  Such behavior is seen in
other doped Mott insulators,\cite{thomas} and systems such as
doped Polyacetylene\cite{heeger} in which solitons are known to
exist.

In this paper, we solve the mean-field equations self-consistently to
determine the internal structure of the soliton and then examine the
motion of the soliton in the antiferromagnetic background.
In the mean-field treatment described below,
the directional fluctuations of the
order parameter are turned off.
Consequently, the mean-field solution
exhibits long-range order which breaks
the lattice translational symmetry.
At low energy and long wavelength the
important dynamical
degrees of freedom are
{\it a)} the motion of the charged
solitons, and {\it b)} the directional fluctuation of
the antiferromagnetic order parameter.
These quantum fluctuations restore
the symmetry broken at the mean-field level.
To obtain an effective action
governing these important dynamical
processes, we force the
antiferromagnetic order parameter to
have both solitons (with time-dependent locations)
and space-time-dependent
directional twists. We then eliminate
the electron degrees of freedom corresponding
to particle-hole pair excitations.

In one dimension, the soliton in the Hubbard model
is intimately related to the structure
soliton in doped Polyacetylene.\cite{su79}
In Polyacetylene, the soliton arises
because the ground state of the undoped
material breaks a two-fold symmetry
of the total electron-phonon Hamiltonian.
For the undoped (i.e. half-filled) Hubbard model in 1D,
there is no known broken symmetry.
The spin-spin correlation
decays algebraically.\cite{affleck}
{}From this point of view it is
surprising that there exist
finite-size solitons upon doping.
The key to the solution of the
puzzle is the realization
of the very different roles played
by the longitudinal (magnitude)
and transverse (direction) components
of the antiferromagnetic order
parameter. A useful picture for the
algebraic order in the 1D Hubbard
model is the following: Imagine
locally (in the sense of both space and time)
the order parameter has a well-developed magnitude.
However, on a much larger
space-time scale, the direction of these
local order parameters fluctuates,
which destroys the long-range order.
In other words, the algebraic
antiferromagnetic correlation is due
to the power-law decay of the
directional correlation between local
order parameters, not due to the absence of
local moments.
In the case of high $T_c$ oxides (2D system),
this picture is
supported by experiments.\cite{neutron}
Thus the
mean-field longitudinal spin order
plays precisely the same role as the
Peierls order does in  Polyacetylene.

The soliton in 1D is a $\pi$-phase-shift domain
wall between the two different
mean-field ground states. In the absence of
doping the soliton is unstable.
In fact once we allow the spin direction to
fluctuate, the soliton decays into an infinite-width
spin twist. Finite-size solitons are stabilized by doping.
As will be shown later, after accommodating a hole, a
charge soliton remains sharp even when the directional
fluctuations are allowed.
The resulting action
governing the charge and spin dynamics
are given by
Eqs.~(\ref{nlsm}) and (\ref{charge_1d}).
Here we note that due to the fact
that the charged solitons are spinless,
the two degrees of freedom are decoupled.

In 2D a domain wall is an extended object. Thus one may
expect that it is more difficult for doping to stabilize it.
For {\it one hole} in a half-filled Hubbard
model with sufficiently big $U/t$ ($t$ is the hopping amplitude and
$U$ is the onsite repulsion strength), we find that this
expectation is indeed true. In this case we find that
the doped charge is accommodated by
a localized {\it non-topological} soliton (or a spin bag).
This soliton is the finite $U$ generalization
of a hole in the $t$--$J$ model:
It has spin quantum number 1/2 and its
spin direction is constrained to
be opposite to the magnetic moment in its
absence.
Here due to the finiteness of $U$,
a small amount of double
occupation exists, and the size of the
soliton depends on $U/t$.
The effective action describing
soliton motion and antiferromagnetic fluctuations
is the $t$--$J$ model.
Due to the coupling to the antiferromagnetic fluctuations,
the soliton will be further dressed to form a quasiparticle.

The quasiparticle formation can be simply understood
in the string picture by considering
the soliton propagation in the antiferromagnetic
background.\cite{dago90,eder94,gan}
As the soliton moves, it leaves behind a string of
overturned spins which
cost antiferromagnetic energy.
This energy cost inhibits the motion of the soliton
and mimics the
effects of a linearly confining
potential.  The bound state of
the soliton in this potential is
the quasiparticle.\cite{gan}  It is important to point
out that spin flipping processes in principle
can erase the string of
frustrated bonds behind the soliton and
enable it  to escape the potential.
If that happens the quasiparticle is no longer
well defined. Exact numerical studies
of one hole in an antiferromagnet, however,
suggest that this does not happen and the
quasiparticle picture
remains valid.
For the 2D Hubbard model with
$t$ and $U$ chosen in the range relevant to the high-$T_c$
superconductors, the quasiparticle
size is about $3 \sim 5$ lattice
spacings.\cite{dago94,dago90,poil93,eder94,haas94,moreo94,%
bulut,germans,trug90,liu92,dago,putz}
We therefore expect that the qualitative properties of
the quasiparticle, such as
its quantum numbers, are not changed by the omitted spin
flipping processes. However, all quantitative aspects are
subject to renormalization.
In addition to their effect on the internal
structure of the quasiparticle,
quantum spin fluctuations have
another important effect -- they
generate coherent propagation of the
quasiparticle as a whole.
The result is that the quasiparticle primarily
hops
between neighboring sites
within the same sublattice.
In the presence of slow
time-dependent twisting of the antiferromagnetic
order, the quasiparticle spin
must follow the direction of
the instantaneous local moment.
Thus
we arrive at the following ``surfing model'':
Imagine  a slowly fluctuating membrane whose
normal directions define the local
directions of the magnetic moments,  then
the quasiparticles surf on the membrane
but have to keep their spins
locally perpendicular to the membrane.
This is illustrated in Fig.~\ref{surf}.
It is known\cite{weigman,shankar,wen,lee} that when a
quantum mechanical particle moves on a
fluctuating curved
surface under the constraint that
its spin must be normal to the local
surface element, it experiences
a Berry phase factor originating from the
overlap of its spin states at different space-time points.
If the curvature of the surface is nonzero,
such a Berry phase cannot be eliminated by
a gauge transformation.
The effect is analogous
to that of a fluctuating magnetic field.
This is precisely the origin of the
fluctuating gauge field in the effective theory,
Eqs.~(\ref{nlsm_2d})~and~(\ref{hqp}),
of the Hubbard (or $t$-$J$) model.

The layout of the rest of the  paper is as follows.
In Sec.~\ref{sect2}, we describe the path integral formalism for the
Hubbard model.  In Sec.~\ref{sect3}, we apply this formalism to the 1D
Hubbard model, show how the standard results are reproduced, and
develop the analogy with Polyacetalene.  In Sec.~\ref{sect4}, we apply
the formalism to the 2D Hubbard model, discuss the connection with
the spin-bag model proposed by Schrieffer, Wen, and
Zhang \cite{schrieffer88}, examine the quasiparticle formation,
and derive an effective action.
In Sec.~\ref{sect5}, we give a conclusion and a summary of our results.

\section{Path integral formalism}  \label{sect2}

The Hamiltonian of the Hubbard model is,
\begin{equation}
H = -t \sum_{<i,j>,\sigma}( c^{\dagger}_{i\sigma}
c_{j\sigma} + c^{\dagger}_{j\sigma} c_{i\sigma} )
+ U \sum_{i} n_{i\uparrow} n_{j\downarrow} ,
\label{hhubb}
\end{equation}
where $c^{\dagger}_{i\sigma}$ is
the electron creation operator and
$n_{i\sigma}= c^{\dagger}_{i\sigma} c_{i\sigma} $.
The summation over $< \!\! i,j \!\! >$
only includes the nearest neighbor
links.
The partition function is given by
\begin{equation}
 Z = \int {\cal D}[c,\bar{c}]   \;
 \exp\left\{ -\int^{\beta}_{0} d\tau \left[ \sum_{i,\sigma}
 \bar{c}_{i\sigma}
 (\partial_{\tau}-\mu) c_{i\sigma} + H \right] \right\} ,
\end{equation}
where $\mu$ is the chemical potential.

The usual way to proceed from here is to
factorize the Hubbard interaction
via a Hubbard-Stratonovich transformation.
There are many ways to factorize the Hubbard
interaction, and in principle they are all
equivalent. However, if one does not perform the
functional integral over the Hubbard-Stratonovich field
exactly, in particular if one only integrates over the
long-wavelength and low-frequency components of the
auxiliary field, then different decoupling scheme gives
different results. The latter is a well known phenomenon.
For example in the mean-field theory (i.e. one only keeps
the $\vec{q}=0,\omega=0$ components of the Hubbard-Stratonovich field)
it is crucial to find the ``right'' decoupling scheme, or equivalently
the right order parameter.
The situation can be even worse.
In some cases the low-$\vec{q}$
and low-$\omega$ components of
the Hubbard-Stratonovich fields
associated with different decoupling scheme
captures different aspect of the low-energy physics.
When that happens one has to include all relevant
auxiliary fields in the effective action. For a prior example
of such multi-auxiliary-field treatment, the reader is referred
to Finkelstein's derivation of the non-linear $\sigma$ model
for the correlated disorder transport problem.\cite{finkelstein}

For the Hubbard model near half-filling, we
expect that the important charge fluctuations are
near momentum $\vec{q}=0$, while the important spin
fluctuations in 2D are around $\vec{q}=(\pi,\pi)$.
In the following we shall
use $\{A\}_{\vec{q}_{0}+\vec{q}}$ to denote
the slow $A$-modes around the modulation
wave vector $\vec{q}_{0}$ and implicitly
impose the condition $|\vec{q} | < \pi$.
Inspired by the above understanding the Hubbard
interaction factorizes into
the following fluctuation
channels,
\begin{eqnarray}
\sum_{j} \bar{c}_{j\uparrow} c_{j\uparrow}
\bar{c}_{j\downarrow} c_{j\downarrow} & \simeq &
\sum_{ |\vec{q}| < \pi } \left[
\frac{1}{4}
\{\bar{c}_{\uparrow} c_{\uparrow}
+ \bar{c}_{\downarrow} c_{\downarrow}\}_{\vec{q}} \;
\{\bar{c}_{\uparrow} c_{\uparrow}
+ \bar{c}_{\downarrow} c_{\downarrow}\}_{-\vec{q}}
- \{\bar{c}_{\uparrow} c_{\downarrow}\}_{\vec{\pi}+\vec{q}} \,
\{\bar{c}_{\downarrow} c_{\uparrow}\}_{-\vec{\pi}-\vec{q}}
	\right.   \nonumber     \\
& & \left.
 - \frac{1}{4}
\{\bar{c}_{\uparrow} c_{\uparrow}
- \bar{c}_{\downarrow}
c_{\downarrow}\}_{\vec{\pi}+\vec{q}} \;
\{\bar{c}_{\uparrow} c_{\uparrow}
- \bar{c}_{\downarrow}
c_{\downarrow}\}_{ -\vec{\pi} -\vec{q}}
	\right] .   \label{factorize}
\end{eqnarray}
where $\vec{\pi}=(\pi,\pi)$.
Note that the factor $1/4$
in front of the first and third terms on the right hand side
of (\ref{factorize}) is obtained by omitting
the following two terms in the factorization:
$(1/4)\{\bar{c}_{\uparrow} c_{\uparrow}
- \bar{c}_{\downarrow} c_{\downarrow}\}_{\vec{q}} \,
\{\bar{c}_{\uparrow} c_{\uparrow}
- \bar{c}_{\downarrow} c_{\downarrow}\}_{-\vec{q}}$
and
$(1/4)\{\bar{c}_{\uparrow} c_{\uparrow}
+ \bar{c}_{\downarrow}
c_{\downarrow}\}_{\vec{\pi}+\vec{q}} \;
\{\bar{c}_{\uparrow} c_{\uparrow}
+ \bar{c}_{\downarrow}
c_{\downarrow}\}_{ -\vec{\pi} -\vec{q}}$.
In the above we have explicitly
omitted the Cooper channel.
The necessary Hubbard-Stratonovich decoupling thus includes
\begin{eqnarray}
U \sum_{j} \bar{c}_{j\uparrow} c_{j\uparrow}
\bar{c}_{j\downarrow} c_{j\downarrow} & \rightarrow &
\sum_{ |\vec{q}| < \pi } \left\{
i \Delta_{c}(\vec{q}) \,
 \{\bar{c}_{\uparrow} c_{\uparrow}
+ \bar{c}_{\downarrow} c_{\downarrow}\}_{-\vec{q}}
+ \frac{1}{U}  \Delta_{c}(\vec{q}) \Delta_{c}(-\vec{q})
	\right. \nonumber       \\
& & + \Delta_{z}(\vec{\pi} +\vec{q}) \,
 \{\bar{c}_{\uparrow} c_{\uparrow}
- \bar{c}_{\downarrow}
 c_{\downarrow}\}_{ -\vec{\pi} -\vec{q}}
+ \frac{1}{U}  \Delta_{z}(\vec{\pi} +\vec{q})
\Delta_{z}(- \vec{\pi} - \vec{q})
	\nonumber       \\      & & \left.
+  \overline{\Delta}_{+}(\vec{\pi} +\vec{q}) \,
\{ \bar{c}_{\uparrow}
c_{\downarrow} \}_{\vec{\pi} +\vec{q}}
+ \{ \bar{c}_{\downarrow}
c_{\uparrow} \}_{\vec{\pi} + \vec{q}} \,
\Delta_{+}(\vec{\pi} +\vec{q})
+ \frac{1}{U} | \Delta_{+}(\vec{\pi} +\vec{q}) |^{2}
 \right\} .
\end{eqnarray}
In coordinate-space we have
\begin{eqnarray}
U \sum_{j} \bar{c}_{j\uparrow} c_{j\uparrow}
\bar{c}_{j\downarrow} c_{j\downarrow} & \rightarrow &
\sum_{ j } \left\{
i \Delta_{c}(j) \,
 ( \bar{c}_{j\uparrow} c_{j\uparrow}
+ \bar{c}_{j\downarrow} c_{j\downarrow} )
+ \frac{1}{U}  \Delta_{c}^{2}(j)
	\right. \nonumber       \\
& & + (-1)^{j} \Delta_{z}(j) \,
( \bar{c}_{j\uparrow} c_{j\uparrow}
- \bar{c}_{j\downarrow}  c_{j\downarrow} )
+ \frac{1}{U}  \Delta_{z}^{2}(j)
	\nonumber       \\      & & \left.
+ (-1)^{j} [ \, \overline{\Delta}_{+}(j) \,
 \bar{c}_{j\uparrow}
c_{j\downarrow}
+  \bar{c}_{j\downarrow}
c_{j\uparrow} \,
\Delta_{+}(j) ]
+ \frac{1}{U} | \Delta_{+}(j) |^{2}
 \right\} .     \label{hs_transf}
\end{eqnarray}
In the above, we have explicitly pulled out the staggering factor from
$\Delta_{z}(j)$ and $\Delta_{+}(j)$.
Therefore, all the remaining variables
$\Delta_{c}(j)$, $ \Delta_{z}(j)$
and $\Delta_{+}(j)$ are  smooth fields.
In 2D, each lattice site
is labelled by two integers,
$j=(j_{x},j_{y})$, in the unit of lattice constant,
and the staggering factor
denotes $(-1)^j=(-1)^{j_{x}+j_{y}}$.

Combining Eqs.~(\ref{hhubb}) to (\ref{hs_transf}), the
partition function can be approximated as,
\begin{eqnarray}
 & & Z \approx \int {\cal D}[c, \bar{c},
\Delta_{c},\Delta_{z},\Delta_{+},
\overline{\Delta}_{+} ]   \;
 \exp\left\{ -\int^{\beta}_{0} d\tau \left[ \sum_{j,\sigma}
 \bar{c}_{j\sigma}
 (\partial_{\tau}-\mu) c_{j\sigma} + H' \right] \right\} ,
\\
 & & H' = -t \sum_{<i,j>,\sigma}
( \bar{c}_{i\sigma}  c_{j\sigma} + h.c. )
+ \frac{1}{U} \sum_{j} \left\{
[\Delta_{c}(j)]^{2} + [\Delta_{z}(j)]^{2}
+ | \Delta_{+}(j)|^{2}
\right\}
	\nonumber       \\
&  &  +  \sum_{j,\alpha\beta}
 \bar{c}_{j\alpha}
\left\{ i \Delta_{c}(j)
+ (-1)^{j} \left[ \Delta_{z}(j) \sigma_{z}
+ \frac{1}{2} \overline{\Delta}_{+}(j)
(\sigma_{x}+i\sigma_{y})
+ \frac{1}{2} \Delta_{+}(j) (\sigma_{x}-i\sigma_{y}) \right]
\right\}_{\alpha\beta}
c_{j\beta} .
	\label{hprime}
\end{eqnarray}
The fact that only smooth configurations of
$\Delta_{c}(j)$,\
$\Delta_{z}(j)$ and
$\Delta_{+}(j)$ are included in the path integral ensures that there is
no overcounting.

To emphasize the SU(2) symmetry we group $\Delta_{z}(j)$ and
$\Delta_{+}(j)$
together and write them as
\begin{equation}
\Delta_{z}(j) \sigma_{z}
+ \frac{1}{2} \overline{\Delta}_{+}(j)
(\sigma_{x}+i\sigma_{y})
+ \frac{1}{2} \Delta_{+}(j) (\sigma_{x}-i\sigma_{y})
=   \Delta_{s}(j) \,  g^{+}(j) \sigma_{z} g(j).
	\label{g_transf}
\end{equation}
Here $\Delta_{s}(j) $ is a real scalar field representing the
longitudinal
component of the local moment, while
$g(j)$ is an SU(2) matrix encoding the transverse components of the
local moment. We
emphasize that both $\Delta_{s}(j) $  and $g(j)$ are smooth functions of space
and time.

Let us now perform a space-time-dependent
local gauge transformation to
absorb the $g$ and $g^+$ in Eq.(\ref{g_transf}).
\begin{equation}
\bar{c}'_{i} = \bar{c}_{i} g^{+}(i),
\hspace{.2in} {\rm with } \; \;
\bar{c}_{i} =(\bar{c}_{i\uparrow},
\bar{c}_{i\downarrow}) .
\end{equation}
We then rewrite everything
in terms of the new variables. A straightforward manipulation shows
that\cite{schulz90}
\begin{eqnarray}
 Z &=&
\int {\cal D}[c,\bar{c}, \Delta_{c},\Delta_{s},g]
 \;  \exp\left\{ -\int^{\beta}_{0}
d\tau {\cal L} \right\}  ,      \\
{\cal L} &=&  \sum_{j} \bar{c}_{j}
 \left[ \partial_{\tau} + g(j) \partial_{\tau} g^{+}(j)
+ (-1)^{j} \Delta_{s}(j)  \sigma_{z}
 + i \Delta_{c}(j) - \mu \right]
c_{j}
\nonumber       \\
& & -t \sum_{<i,j>}
\left[ \bar{c}_{i} g(i)g^{+}(j)
c_{j} + h.c. \right]
+ \frac{1}{U} \sum_{j}
\left[ \Delta_{c}^{2}(j)+ \Delta_{s}^{2}(j)
\right] .
	\label{laglocal}
\end{eqnarray}
In the above we have omitted the primes on the Grassmann variables. Our
subsequent treatments of the path integral~(\ref{laglocal}) involve a saddle
point analysis followed by expansions that include small fluctuations around
the
latter. In specific our saddle analysis assumes
time-independent (but spatially dependent) fields given by
\begin{eqnarray}
&\Delta_{c}(\tau,j) = - i  \Delta_{c}^{(0)}(j) ,
	\label{sads1} \\
&\Delta_s(\tau,j)=\Delta_s^{(0)}(j), \\
&g(\tau,j)=g_0(j),  \label{sads3}
\end{eqnarray}
and the corresponding saddle point equations are
\begin{eqnarray}
& &\Delta_{c}^{(0)}(j) =
\frac{U}{2} \langle n_{j} \rangle , \\
& &\Delta_{s}^{(0)}(j) {\rm Tr} \left[ g_{0}^{+}(j)\sigma_z g_{0}(j)
\sigma_{\alpha} \right]
= - U (-1)^{j}  \langle  c^{\dagger}_{j} \sigma_{\alpha}
c_{j} \rangle ,  \hspace{.2in} \alpha=x,y,z.
\end{eqnarray}
In the above $n_{j}\equiv\sum_{\sigma} c^{\dagger}_{j\sigma}
c_{j\sigma}$.
This analysis is equivalent to the Hartree-Fock solution.
It is also equivalent to the variational solution constructed using a
Slater
determinant as the
trial wave function.

\section{The one-dimensional Hubbard model revisited} \label{sect3}

The purpose of this section is to demonstrate that our approach
reproduces the
well known results in 1D.
Hopefully it will strengthen the readers' confidence (as well as ours)
when it
is applied to 2D.

\subsection{1D Hubbard model at half-filling: the
non-linear $\sigma$ model} \label{halffill}

To warm up, let us first consider the 1D Hubbard model at
half-filling (i.e. with on average one electron per site). It is known
that for
arbitrary repulsion, the half-filled Hubbard model in 1D is a charge
insulator,
and  the only low-energy excitations are spin in nature. Here we derive
the
effective theory for such spin excitations.

We first find the saddle point solution in the form of
Eqs.(\ref{sads1})--(\ref{sads3}). The result is
\begin{equation}
	\Delta_{c}(j)=- i \Delta_{c}^{(0)},
\hspace{.3in} \Delta_{s}(j) = \Delta_{s}^{(0)}  ,
\hspace{.3in} g(j)=1 .
\end{equation}
The corresponding
saddle point(or mean field) Hamiltonian is
\begin{equation}
H_{mf} =  -t \sum_{<i,j>}
\left( c^{\dagger}_{i}
c_{j} + h.c. \right)
+ \sum_{i} c^{\dagger}_{i}
\left[  (-1)^{i}  \Delta_{s}^{(0)}   \sigma_{z}
- \widetilde{\mu} \right] c_{i}
+\frac{1}{U} \left[ (\Delta_{s}^{(0)} )^{2}
  - (\Delta_{c}^{(0)})^{2}
\right]   {\cal N} ,
	\label{hmf}
\end{equation}
where ${\cal N}$ is the number of lattice sites and
\begin{equation}
\widetilde{\mu} = \mu - \Delta_{c}^{(0)} .
\end{equation}
This Hamiltonian can be easily diagonalized. At half-filling
$\widetilde{\mu}=0$
and $\Delta_{s}^{(0)}$ satisfies\cite{note}
\begin{equation}
\frac{2}{U} =  \int^{2\pi}_{0} \frac{dk}{2\pi}
\frac{1}{\sqrt{ (2t \cos k)^{2}
+ \left( \Delta_{s}^{(0)} \right)^{2} } } .
	\label{sol_1d}
\end{equation}
	%This Eq. gives a gap proportional
	%to $\exp(-2 \pi t /U)$.
For $U/t=8/3, \, 4$, and $8$, the solutions to Eq.(\ref{sol_1d}) give
$ \Delta_{s}^{(0)}\approx 0.726t$, $1.539t$, and $3.754t$
respectively.
The solution describes a long-range ordered antiferromagnets with the
local
moment $m_0=0.272, \; 0.384$ and $0.469$  respectively.
The mean-field electronic spectrum has a gap separating an empty
conduction band
from a full valence one.

To study the low-energy excitations, we include fluctuations in
$\Delta_{c}(j)$,\  $ \Delta_{s}(j)$
and $g(j)$ around their saddle point values, and integrate out $c$ and
$\bar{c}$. Since the action is quadratic in the
fermion variables after Hubbard-Stratonovich
factorization, the integration produces a
fermion determinant. It turns out that both $\Delta_{c}(j)$ and
$ \Delta_{s}(j)$ fluctuations are gapful,
thus can be safely ignored at low energies.
We expand the fermion determinant to quadratic order
in $g\partial_{\mu}g^{+}$ to obtain (the details are supplied in
appendix B)
\begin{equation}
{\cal L}_{\rm eff} =  \sum_{i}  \langle S_{i}^{z} \rangle
{\rm Tr} \left[ \sigma_{z} g(i) \partial_{\tau} g^{+}(i)
\right]
+ \frac{ \rho_{s} }{2}
\int dx \sum_{\mu=v_{F}\tau,x} \left[
{\rm Tr} \left( \partial_{\mu} g \partial_{\mu} g^{+}
\right)
+ {\rm Tr} \left( \sigma_{z} g \partial_{\mu} g^{+}
\sigma_{z} g \partial_{\mu} g^{+} \right) \right].
\label{leff0}
\end{equation}
In the above, $\langle S_{i}^{z} \rangle =
\langle c^{\dagger}_{i} \sigma_{z} c_{i}
\rangle /2 = (-1)^{i} m_0$ is the mean
field magnetization,
and $\rho_{s}$ is the spin stiffness
given by Eq.~(\ref{stiff_s1}).
The spin wave velocity turns out to
be the same as the bare Fermi velocity.
The second term in (\ref{leff0})
is the well known non-linear
$\sigma$ model
written in terms of the SU(2) matrix $g(i)$.
Using $\vec{n} \cdot \vec{\sigma} = g^{+} \sigma_{z} g$,
one can easily verify that
\[ (\partial_{\mu} \vec{n} )^{2} =
{\rm Tr} \left( \partial_{\mu} g \partial_{\mu} g^{+}
\right)
+ {\rm Tr} \left( \sigma_{z}  \partial_{\mu} g^{+}
\sigma_{z} g \partial_{\mu} g^{+} \right).  \]
The first term of (\ref{leff0}) represents a Berry phase.
To see this, we express
\begin{equation}
g= \left( \begin{array}{cc}
		z_{1}^{*}  &  z_{2}^{*}  \\
		-z_{2} &  z_{1}  \\
	\end{array} \right),
		\label{g_rep}
\end{equation}
where $\sum_{\sigma} z^{*}_{\sigma} z_{\sigma}=1$.
In terms of  $z^{*}=(z_{1}^{*}, z_{2}^{*})$,
$\vec{n} = z^{*} \vec{\sigma} z$. It is straightforward to
show that
\begin{eqnarray}
m_0 \sum_{i} (-1)^{i}
{\rm Tr} \left[ \sigma_{z} g(i) \partial_{\tau} g^{+}(i)
\right]
&=& 2m_0  \sum_{i} (-1)^{i} z^{*}(i)
\partial_{\tau}  z(i)
	\nonumber	\\
&=& i \frac{\theta}{2\pi} \int dx \; \partial_{x }
\vec{n} \cdot
( \vec{n} \times  \partial_{\tau} \vec{n} ) ,
\hspace{.2in}  \theta=2\pi m_0.
		\label{berry}
\end{eqnarray}
While our approximate calculations do yield the right form for the
effective
action, it makes a serious mistake. It is well known that the $\theta$
in the effective action is quantized to have the value
$\pi$.\cite{affleck} Our approximate calculation gives a
$\theta$ that depends on the magnitude of the local moment.
It is also known that
the low-energy behavior of the non-linear sigma model critically depends
on the value of $\theta$.\cite{haldane,pruskin}  In particular,
the spin excitation spectrum is gapless
only when $\theta=(2n+1)\pi$. Since $m_0\le 1/2$, we would
conclude that except for the case of $U=\infty$ (in that case
$m_0\rightarrow
1/2$) the spin spectrum is always gapful. This is, of course,
incorrect.
Then what is wrong with
our derivation?

The answer relies on the fact that the value of $\theta$ reveals a
symmetry
of the problem. The original 1D Hubbard model is invariant under the
translation by one lattice spacing. Since it is known that the
ground state of the model preserves this symmetry, it better be true
that
any low energy effective theory is revealing it. This is not so in
Eq.~(\ref{berry})
unless $\theta={\rm integer} \times \pi$.
Indeed, as $x \rightarrow x+a$, $m_0\rightarrow -m_0$
which invert the sign of $\theta$. Since $\frac{1}{2\pi} \int dx \;
\partial_{x }
\vec{n} \cdot ( \vec{n} \times  \partial_{\tau} \vec{n} )$ is an
integer,
the latter is a symmetry only when $\theta$ is an integer multiples of
$\pi$.

Thus if we did the calculation exactly, $\theta$ should turn out to be
$\pi$.
This has been shown to be the case recently by
Nagaosa and Oshikawa~\cite{nagaosa}.
The point is that approximations are not allowed in
extracting quantized quantities.
To proceed, we offer the following symmetry based
arguments to fix the value of $\theta$.\cite{affleck}
 Since from translational symmetry
$ \theta={\rm integer}\times \pi$, our job is only to determine which
integer. For this purpose
we recall that as a function of $U/t$ the half-filled Hubbard model
experiences no phase transition.
Thus it is sufficient for us to determine $\theta$ in the
$U/t\rightarrow\infty$ limit. Since in the latter limit the
leading order result becomes exact, and since $\theta$ remains
quantized as we vary $U/t$, we can deduce its value exactly.
As $U/t\rightarrow\infty$, $m_0\rightarrow 1/2$ and
$\theta\rightarrow\pi$, thus we conclude that the exact value of
$\theta$ in
Eq.~(\ref{berry}) is $\pi$.

Summarizing the above discussions, we have arrived at the following
low-energy
effective theory for the 1D Hubbard model at half-filling:
\begin{equation}
S_{\rm eff} = \int^{\beta}_{0} d\tau \int dx \; \left\{
\frac{i}{2} \; \partial_{x } \vec{n} \cdot
( \vec{n} \times  \partial_{\tau} \vec{n} )
+ \frac{ \rho_{s} }{2}
\left[ \frac{1}{v_{F}^{2}}
( \partial_{\tau} \vec{n} )^{2}
+  (\partial_{x}
\vec{n} )^{2}
\right]  \right\}.      \label{nlsm}
\end{equation}
The consequence of this action is that
transverse  fluctuations  destroy the
long-range order in the saddle point solution. The result is a
quasi-long-range antiferromagnetically
correlated state in which the correlation between
local N\'{e}el order decays as the inverse of distance.

\subsection{The charge and spin
excitations away from half-filling}
\label{h_spinon}

In this section, we shall show that away from half-filling
the elementary excitations of the 1D Hubbard model
are holons and spinons. In addition we shall demonstrate the
similarities between the holon and the soliton in
Polyacetylene.

Our starting point is the saddle-point solution of Eq.~(\ref{laglocal}).
Naively one might take the half-filled solution and take out an electron
from
the valence band. This is dangerous since one overlooks the effect of
doping on
the saddle-point solution. For example in the presence of a static hole,
we do
not expect the saddle point solution to have spatially
uniform spin and charge density.
This is because the spin and charge densities are
self-consistently determined by the occupation of
the electronic states. Around the hole,
the electron density is reduced so that the magnitude of
$\langle S_{i}^{z} \rangle $ must also be reduced.
Motivated by the above considerations we look for a self-consistent
solution of
the following form:
\begin{equation}
\Delta_{c}(j)=-i\Delta_{c}^{(0)}(j),
\hspace{.2in}
\Delta_{s}(j)=\Delta_{s}^{(0)}(j),
\hspace{.2in} g(j)=1.   \label{sadd_point}
\end{equation}
The saddle point Hamiltonian is
\begin{eqnarray}
H_{mf} &=&  -t \sum_{<i,j>}
\left( c^{\dagger}_{i}
c_{j} + h.c. \right)
+ \sum_{j} c^{\dagger}_{j}
\left[ \Delta_{c}^{(0)}(j) - \mu
+  (-1)^{j} \Delta_{s}^{(0)}(j) \;
\sigma_{z} \right] c_{j}
	\nonumber	\\
& & + \frac{1}{U} \sum_{j}
\left\{   \left[ \Delta_{s}^{(0)}(j) \right]^{2}
- \left[ \Delta_{c}^{(0)}(j)\right]^{2}  \right\}.
	\label{hmfkink}
\end{eqnarray}
In Appendix~\ref{app_a}, we show that the continuum
version of Eq.~(\ref{hmfkink})
is very similar to the
corresponding continuum
Hamiltonian for Polyacetylene.\cite{tlm}
The saddle point equations are
\begin{eqnarray}
\Delta_{s}^{(0)}(i) &=& -(-1)^{i} \frac{U}{2} \left[
\langle c^{\dagger}_{i\uparrow} c_{i\uparrow} \rangle
- \langle c^{\dagger}_{i\downarrow} c_{i\downarrow} \rangle
\right] ,
		\label{consis}  \\
\Delta_{c}^{(0)}(i) &=&  \frac{U}{2} \left[
\langle c^{\dagger}_{i\uparrow} c_{i\uparrow} \rangle
 + \langle c^{\dagger}_{i\downarrow} c_{i\downarrow} \rangle
\right] .       \label{consis_c}
\end{eqnarray}

We solve the above equations numerically in
a finite chain of 100 sites with 99
electrons under the open boundary condition.
Armed with the experience from
Polyacetylene,
we choose  the starting point of the self-consistent iteration
to be
\begin{equation}
\Delta_{s}^{(0)}(i)= \Delta_{s}^{(0)}
\tanh\left(\frac{i-x_{0}}{\xi}\right),
\hspace{.2in}
\Delta_{c}^{(0)}(i)=\frac{U}{2},       \label{init_1d}
\end{equation}
where $x_{0}$ and $\xi$ are two arbitrary
parameters
and $ \Delta_{s}^{(0)} $ is the solution at half filling.
It turns out that for small $U/t$,
the self-consistent iteration preserves $x_{0}$.
The resulting self-consistent saddle point
solution suggests that $x_{0}$ does not necessarily center at a high
symmetry position (such
as a lattice site or the mid-point of a lattice bond).
To ensure that the 100-site chain is sufficiently long,\cite{kivelson}
we have checked that the difference between the values of
$\Delta_{s}^{(0)}$
under periodic and open boundary conditions at half filling only shows
up at the fourth significant figures.
For $U/t=8/3$ the saddle point solution and the associated spin
and charge profiles with
the kink centered at the middle of a lattice link are shown
in Fig.~\ref{var_kink_e} and Fig.~\ref{kink_profile} respectively.

We notice that the boundary effects
die off a few lattice spacings away
from the two ends of the chain.\cite{kivelson}
The ground state energy at the saddle point
is $-0.196t \, (-0.269\Delta_{s}^{(0)})$ for $U/t=8/3$, where the
reference energy is chosen to be the energy
of the ``doped semiconductor state'' obtained by
removing one electron from the top of the valence band of the
half-filled self-consistent band structure under
open boundary condition.
For other values of $U/t$ ratios, the energies
of the saddle point solutions with $x_{0}$ situated at the mid-point
of a bond are
$-0.242t \, (-0.260 \Delta_{s}^{(0)})$ for $U/t=3$,
$-0.366t \, (-0.238 \Delta_{s}^{(0)})$ for $U/t=4$,
and $-0.643 t\, (-0.171 \Delta_{s}^{(0)})$ for $U/t=8$.

The one-electron spectrum for
$U/t=8/3$
is shown in Fig.~\ref{kink_spec}.
As in Polyacetylene, there are two midgap
states. These states have opposite spin quantum numbers, and are almost
degenerate.
Their wave functions are shown in Fig.~\ref{loc_wavefunc}.
All energy levels below these
two states are occupied.
The spin and charge densities of the
saddle point solution can be fitted
to simple analytic functions.
The least-square fitting gives, for $U/t=8/3$,
\begin{eqnarray}
& & (-1)^{i} \langle S_{i}^{z} \rangle
= \frac{\Delta_{s}^{(0)}}{U} \tanh\left(
\frac{i-x_{0}}{\xi_{s}}\right) ,
\hspace{.2in}  \xi_{s}=2.978,    \label{spind_1d} \\
& & \langle n_{i} \rangle
= 1 - \frac{1}{2 \xi_{c} \cosh^{2}\left(
\frac{i-x_{0}}{\xi_{c}}\right) },
\hspace{.2in} \xi_{c}=3.209 .     \label{charged_1d}
\end{eqnarray}
The quality of the fitting is demonstrated
in Fig.~\ref{verif_consis}.
In principle, we still have to show that we have allowed sufficient
variational degrees of freedom in our mean-field
ansatz Eq.(\ref{sadd_point}). Here we skip this
issue by referring the reader to the similar situations
in Polyacetylene where the
the soliton stability was proven.\cite{heeger}

Intuitively one expects that the kink energy reaches
optimum if the center of the kink is located at high symmetry
positions of the chain.
It turns out that for e.g.
$U/t=8/3$ the kink energy
has no measurable dependence on its position!
This is, of course, what one expects
based
on the continuum Hamiltonian we derive
in Appendix~\ref{app_a}. The validity of the continuum approximation is
that
the self-consistent spin and charge densities have smooth spatial
variations. This requirement will eventually
be violated when $U/t\gg 1$, in
which case the kink is a very localized object.
In the latter case the center of the kink becomes a
crucial parameter, not only does it affect the energetics but it also
determines
whether self-consistency can be reached at all. Indeed, we
have found
that for large $U/t$ the only possible self-consistent kink solution
is a kink centered at the mid-point of a bond.

The static kink solution discussed above represents a
snap shot of the charged quasiparticle of the 1D
Hubbard model. If the time scale associated with the motion of
the quasiparticle is long compared with the inverse excitation gap of the
soliton band structure, to a good approximation we can regard it as a rigid
boost of the static soliton discussed above. In that limit, in order
to determine the quantum numbers of a moving soliton it is
sufficient to study the corresponding quantities assuming the soliton is
at rest. In the following we shall assume such an adiabatic picture. Let us
first determine the charge of a kink. Since we are interested in the case where
a continuum description is possible, we shall concentrate on the
case of $U/t=8/3$. Moreover we shall assume that the center of the
soliton
is situated at the center of a bond. In Fig.~\ref{verif_consis} we show
the change in site occupation number due to the presence of a kink.
{}From Eq.~(\ref{charged_1d}) we deduce that
\begin{equation}
\sum_{i} \langle \delta n_{i} \rangle
\simeq - \int^{\infty}_{-\infty} dx \frac{1}{2\xi_{c}
\cosh^{2}\left(\frac{x-x_{0}}{\xi_{c}}\right)}=-1.
\end{equation}
Therefore, the kink carries charge $e$.

Now we determine the total $S^{z}$ associated with a kink.
This is sufficient for the purpose of later discussions
because it turns out that $S^z$ plays the role
of coupling constant in the residual
gauge coupling between the soliton current and the
fluctuating antiferromagnetic background (see later).
The total $\langle S^z\rangle$ associated with a
kink is given by,
$\langle S^z\rangle=\sum_{i} \langle S_{\rm pair}^z(i) \rangle$, where
$ \langle S_{\rm pair}^z(i) \rangle
\equiv \langle S^z(i)+S^z(i+1) \rangle/2$.
In Fig.~\ref{ave_profile}
we show the $\langle S_{\rm pair}^{z}(i) \rangle $
profile, where we
see that the only nonzero $\langle S_{\rm pair}^{z}(i) \rangle $ appears
near the kink.(The zigzag pattern near the chain ends is caused by the
boundary and it is not shown in the figure.)
Had we plotted the same quantity for the half-filled case we would have
found zero everywhere except near the ends of the chain.
{}From Fig.~\ref{ave_profile},
It is clear that by summing up all
$\langle S_{\rm pair}^{z}(i) \rangle $ we get zero.

Thus on a scale larger than the size (determined by the
larger one between $\xi_{c}$ and $\xi_{s}$ ) of the kink
we can regard it as a point object carrying charge $e$ and
zero $ S^{z}$.
These quantum numbers are consistent with those of the
holon in the 1D Hubbard model.

Next we turn to the neutral spin excitations.
In the presence of the longitudinal spin order
one naturally expect the low-lying spin excitations
to be described by the smooth space-time fluctuating $g(\tau,i)$.
However, is that all? For instance
can we have a neutral
spin excitations with a sharp kink
form in  $\Delta_{s}^{(0)}(i)$ ?
In the following
we show that the answer is no.
Specifically, we shall start from the
solution corresponding to a charged soliton and
show that if the missing electron is put back into
the system the soliton decays into a smooth spin twist.
For this purpose, it suffices to
find a path connecting the configurations corresponding to the kink
and the smooth twist, and show that the
energy continuously decreases along the path.
Considering Eqs.~(\ref{hprime}) and (\ref{g_transf}),
the path we found is,
\begin{eqnarray}
& &\Delta_{c}^{(0)}(i) + (-1)^{i}
 \Delta_{s}^{(0)}(i) g^{+}(i) \sigma_{z} g(i) =
1 - \frac{1-\lambda}{2 \xi_{c}  \cosh^{2}\left(
\frac{i-x_{0}}{\xi_{c}}\right) }
	\nonumber	\\	& & \hspace{.4in}
+ (-1)^{i} \Delta_{s}^{(0)} \left[
\tanh\left(\frac{i-x_{0}}{\xi_{s}}\right) \sigma_{z}
+ \frac{\lambda}{\cosh\left(
\frac{i-x_{0}}{\xi_{s}}\right) } \sigma_{x} \right].
	\label{intro_lambda}
\end{eqnarray}
The first segment of the path
is given by increasing $\lambda$ from zero to one
while keeping $\xi_{c}$ and $\xi_{s}$ fixed.
We note that when $\lambda=1$, the
spin order  profile
given by Eq.~(\ref{intro_lambda})
becomes a twist of the size $\xi_{s}$.
The second part of the path is
to increase $\xi_{s}$ while fixing $\lambda=1$.
Again we choose a chain of 100 sites with $U/t=8/3$ and
$x_{0}$ situated at the center of a bond
for the demonstration. To obtain the electron
wave functions along the path we modify the
Hamiltonian~(\ref{hmfkink}), by replacing
the onsite potential term with the
right hand side of Eq.~(\ref{intro_lambda}).
We diagonalize this Hamiltonian to
construct the wave function by filling 100 electrons in the lowest
100 energy levels. Then we obtain the variational energy
by calculating the expectation value
of the full Hubbard Hamiltonian. The result is shown in
Fig.~\ref{var_xi_lamb}, where the the reference energy is
chosen to be the energy expectation of the  half-filled
mean-field state.
In the left panel, the energy is shown
as the function of $\lambda$ while
fixing $\xi_{s}=2.978$ and $\xi_{c}=3.209$.
We see that the kink is unstable once
an additional electron is introduced.
This is not difficult to understand if we
recall the physics of Jahn-Teller effect.
The two nearly degenerate midgap states
are empty in the case of charged soliton.
If the missing electron is put back,
it is energetically favorable to
split the two midgap levels
and occupy the lower level.
The off-diagonal $\sigma_{x}$ part
in Eq.~(\ref{intro_lambda})
precisely provides matrix element between
these two states and consequently split them.
In the right panel of
Fig.~\ref{var_xi_lamb},
the variational energy is shown
as a function of the twist size $\xi_{s}$
while fixing  $\lambda=1$.
The twist given by Eq.~(\ref{intro_lambda})
with $\lambda=1$
further decays to an infinitely smooth one,
as illustrated in Fig.~\ref{spinon_profile}.
In fact, the twist is just a spinon.\cite{emery}

\subsection{Effective theory}           \label{effec_1d}

In the above discussions the saddle point solution
for doped 1D Hubbard model consists of static and
localized solitons (each accommodating one hole)
separating the otherwise perfectly ordered
antiferromagnetic domains. Ignored
in this mean-field theory are the smooth space-time variation
of the direction of the order parameter, and the
propagation the soliton
configurations. In general it is not clear
that the saddle point solution in the
presence of $N$ holes is $N$ separate
solitons (for example, it is important
to compare the mean-field energies
associated with, say, a phase-separated
solution with that of $N$ equally spaced solitons).
In the rest of this section we will assume
that the individual soliton remains
stable against finite doping. Under that
assumption we will show that the low-energy
theory is precisely the effective theory
for the Hubbard model obtained via non-Abelian bosonization
(see Appendix~\ref{cal_nlsm}).

We now briefly describe
the derivation of the effective action,
and leave the details in Appendix~\ref{cal_nlsm}.
The basic strategy is to allow the solitons
to have time-dependent locations,
and twist the spin directions in a space-time dependent way.
Then we evaluate the cost in action due
to such distortions.

At the length scale longer than the
kink size $\xi_{s}$, the kink (\ref{spind_1d})
can be simply viewed as a point at which the
longitudinal spin order parameter $\Delta_{s}(i)$ flips its sign.
Consequently at a given site the sign
of $\Delta_{s}(i)$
has changed
as many times as the number of holons
to its left:
\begin{equation}
 \Delta_{s}(x) \simeq \Delta_{s}^{(0)}
 \; \cos \left[ \pi
\int_{0}^{x} dy \, n(y) \right] , \hspace{.2in}
0\leq x \leq L,
\end{equation}
where
\begin{equation}
n(y) =
\sum_{i=1}^{N_{h}} \delta(y-x_{i}) ,
\end{equation}
is the holon density. At low energy and long wavelength
we expect the holon density to be its average value $n_{h}$
plus small perturbations:
$n(x)=n_{h} + \delta n(x)$.
The spin order profile
in the finite doping becomes, upon
replacing the lattice site $j$ with
the continuum coordinate $x$,
\begin{eqnarray}
& & (-1)^{j} \Delta_{s}(j) \rightarrow \Delta_{s}^{(0)}
\cos \left[ 2 k_{F} x + \Phi(x) \right] ,
	\label{intro_Phi}       \\
& & \Phi(x) = - \pi \int_{0}^{x} dy \; \delta n(y) .
	\label{slide_phase}
\end{eqnarray}
where $k_{F}=\pi (1-n_h)/2$.
Furthermore, the phase $\Phi(x)$
satisfies the following density-phase conjugation relation:
\begin{equation}
\frac{\partial \Phi(x) }{\partial x} =
- \pi \; \delta n(x) .
	\label{den_phas}
\end{equation}
Substituting Eq.~(\ref{intro_Phi})
into Eq.~(\ref{laglocal}), and integrating out $c$ and $\bar{c}$,
we derive the desired low-energy
effective theory in the case of finite doping.
The charge part is given by
a free boson theory
\begin{equation}
S_{\rm eff}^{(c)} =
\frac{ \rho_{c} }{2}
\int_{0}^{\beta} d\tau \int dx
\left[ \frac{1}{  v_{c}^{2} }
(\partial_{\tau} \Phi)^{2}
+ (\partial_{x} \Phi)^{2} \right] ,
	\label{charge_1d}
\end{equation}
where $\rho_{c}$ and $v_{c}$
are the charge stiffness and velocity,
given by Eqs.~(\ref{stiff_c1})
and (\ref{velo_c1}).
The spin part is still given by
Eq.~(\ref{nlsm}).
Both $\rho_{c}$ and $v_{c}$
are  different
from $\rho_{s}$ and $v_{F}$.
The detailed calculation is included
in Appendix~\ref{cal_nlsm},
(also see Reference~\onlinecite{nagaosa}.)

As noted by Lee from the
Bethe Ansatz solution\cite{lee90},
the Fermi velocity
in the low-energy effective
action is not much different
from the bare one
of the non-interacting case.
This is consistent with the fact that
in Appendix~\ref{cal_nlsm}
the integral in Eq.~(\ref{stiff_s1})
is independent of $ \Delta_{s}^{(0)}  $.
We note that the parameters in the
effective action are subject
to the renormalizations
due to short distance fluctuations
which are simply dropped
when we take the continuum limit
in Appendix~\ref{cal_nlsm}.
However, as the order of
magnitude is concerned,
the results are reliable.
This is in sharp contrast
to the situation in two dimensions
where the effective Fermi velocity
is renormalized down to
the order of $J$($\sim t^{2}/U$), as we shall
see below.

\section{The Hubbard model in two dimensions} \label{sect4}

Since the discovery of the high temperature superconductors,
Anderson\cite{ande87} has been suggesting that the one-band
Hubbard model in two dimensions
(or the closely related $t$--$J$ model\cite{zhang88}) captures
the essential
physics of the CuO$_{2}$ plane in the cuprate
superconductors.
Ever since then, the properties of 2D Hubbard model (on square
lattice in particular)
has been a focus of investigations.
An important parameter for the cuprates is the doping level.
For example, within a class of compounds, the samples with the highest
$T_c$ usually have doping level around 15\%. In terms of the
Hubbard model it means that the averaged site occupation number is
around 0.85. As in one dimension the starting point of our discussion is the
half-filled Hubbard model.

By now it is a common consensus that at half-filling,
the Hubbard model describes an ordered
antiferromagnetic Mott insulator
and  the long wavelength effective theory is a
2+1 D non-linear $\sigma$ model\cite{chak89}.
The derivation of the latter
from the Hubbard model
proceeds in the same way as
in one dimension except that
the Berry phase term
is canceled out in two dimensions.\cite{berry}
The antiferromagnetic order present in the half-filled
2D Hubbard model reflects the fact that the non-linear
$\sigma$ model can be ordered in 2+1 dimensions.

An important question is how to proceed when the
antiferromagnet is doped.
In the conventional approach,
one  performs the Schrieffer-Wolf
transformation on the Hubbard
model to eliminate states having
doubly occupied sites
in the  $U/t \rightarrow \infty$ limit.
The result is the familiar $t$--$J$ model.
In the following we shall adopt
the same approach as the one used in the last section.
As we shall see, there are two levels of quasiparticle
formation. At the first level, like in 1D, we shall find that in
the mean-field theory doping in the antiferromagnetic long-range
ordered state produces solitons, whose corresponding
band structure exhibits states inside the gap. Unlike 1D, the solitons
are non-topological. At the second level, due to the residual coupling
between the soliton and the spin degrees of freedom, the soliton is
further dressed so that the bandwidth associated with its hopping is reduced
from $\sim t$ to $\sim J$. In 1D the second level of renormalization
is absent.

\subsection{Mean-field solution for one hole}

The Hubbard-Stratonovich decoupling scheme and the subsequent mean-field
ansatz are the same as in 1D. In particular, the mean-field
Hamiltonian is
\begin{eqnarray}
H_{mf} &=&  -t \sum_{<i,j>}
\left( c^{\dagger}_{i}
c_{j} + h.c. \right)
+ \sum_{j} c^{\dagger}_{j}
\left[ \Delta_{c}^{(0)}(j)
+  (-1)^{j_{x}+j_{y}} \Delta_{s}^{(0)}(j) \;
\sigma_{z} \right] c_{j}
	\nonumber	\\	& &
+ \frac{1}{U} \sum_{j}
\left\{   \left[ \Delta_{s}^{(0)}(j) \right]^{2}
- \left[ \Delta_{c}^{(0)}(j)\right]^{2}  \right\}.
	\label{hmf_2d}
\end{eqnarray}
The only difference is that now $i,j$ label the square lattice sites.
The corresponding saddle point equations are similar
to Eq.~(\ref{consis}) and (\ref{consis_c}),
\begin{eqnarray}
\Delta_{s}^{(0)}(i) &=& -(-1)^{i_{x}+i_{y}} \frac{U}{2} \left[
\langle c^{\dagger}_{i\uparrow} c_{i\uparrow} \rangle
- \langle c^{\dagger}_{i\downarrow} c_{i\downarrow} \rangle
\right].
		\label{consis1}         \\
\Delta_{c}^{(0)}(i) &=& \frac{U}{2} \left[
\langle c^{\dagger}_{i\uparrow} c_{i\uparrow} \rangle
+ \langle c^{\dagger}_{i\downarrow} c_{i\downarrow} \rangle
\right].  	\label{sdeq_c_2d}
 \end{eqnarray}
We solve the above equations
numerically
on a $14 \times 14$ lattice
with periodic boundary condition.
The number of electrons is such that there is only one hole.
We have chosen three different starting points
for the self-consistent iteration. In each case, we
reduce the N\'{e}el order parameter from its value at half-filling
symmetrically around a
lattice site, lattice link, and the center of a plaquette respectively.
For
$U/t$ not too small (say $U/t\geq 10/3$), the first two starting points
converge to a non-topological soliton
centered on a lattice site. We have also checked that a small perturbation in
the starting configuration does not affect the final self-consistent
solution. The magnitude of the resulting order parameter
has a fourfold rotational symmetry about the center.
For $U/t \geq 10/3$, the third starting point first converges to
a diagonal cigar-shaped non-topological soliton
situated at the plaquette center.
However, if we slightly break the reflection symmetry
about the plaquette center
along the ridge of the diagonal cigar-shaped spin profile,
the iteration further converges to
the same solution as found from
the previous two starting points.
For $U/t \geq 10/3$, the self-consistent
mean-field solutions are non-topological solitons.
The typical
spin and charge density profiles are
shown in Fig.~\ref{bag_profile}, and the corresponding mean-field
band structure is
shown in Fig.~\ref{bag_levels}.
We note that two midgap
states are present. For one hole, only states in
the lower band are occupied.
The energies of the soliton with respect to a hole
at the top of the valence band of the half-filled band structure
are $\delta E=-0.051t \,(-0.049 \Delta_{s}^{(0)} )$
for $U/t=10/3$,
$-0.386t\,(-0.155  \Delta_{s}^{(0)})$ for $U/t=6$,
  %This energy is -0.38554 in 14X14 laatice
  % and -38516 in 10X10 lattice.
$-0.650t \, (-0.182 \Delta_{s}^{(0)}  )$ for $U/t=8$,
$-0.858t\,(-0.185 \Delta_{s}^{(0)} )$ for $U/t=10$,
and $ -1.017t \,(-0.179 \Delta_{s}^{(0)} )$  for $U/t=12$.
For $U/t \geq 8$, the calculations are
done in $10\times 10 $ lattice since the finite
size effect is small for large $U/t$ ratios.

In Fig.~\ref{bag_change},  we show the changes in the spin
density induced by a soliton.
%The local changes of the magnetization
%from  half-filling
%are  shown in Fig.~\ref{bag_change}.
We have checked numerically that
$\sum_{i,\sigma} \delta
\langle c^{\dagger}_{i\sigma}
c_{i\sigma} \rangle = -1$  and $ | \sum_{i} \delta
\langle S^{z}_{i} \rangle | = 1/2$.
Thus, unlike 1D, the soliton carries both charge and spin.

The same mean-field studies for one hole in the
2D Hubbard model
has been carried out by Su and Chen,\cite{su88} and
Choi and Mele.\cite{mele88}
For large $U/t$
our results are consistent with
the their findings. For example,
we have found that
for $U/t \geq 10/3$,
the saddle point solution is a soliton
whose associated spatial variation of
the N\'{e}el order parameter respects the fourfold
rotational symmetry of the underlying lattice. Moreover,
our self-consistent band structure agrees with
that found in Ref.\onlinecite{su88} (for $U/t=5$ on $10\times 10$
lattice)
within numerical uncertainties.
However, for smaller $U/t$ (say $U/t=2$)
we found that unlike
previous claims that the solitons
are cigar-shaped,
after a large number of self-consistent iterations the soliton
converges to a linear bag running diagonally across the whole
finite lattice. Based on the length scale defined by
$v_{F}/2\Delta_{s}^{(0)}$ we
expect significant finite size effects
for small $U/t$. In addition,
for small $U/t$  the detailed structure
of the non-interacting band
(such as the magnitude of the next nearest
neighbor hopping) starts to have a stronger effect on the
soliton shape. At present we have not studied enough lattice sizes
and different boundary condition to
establish that the extended soliton is the genuine solution in the
thermodynamic limit.
In that context it is interesting
to observe that Schulz\cite{schulz} suggested that at finite hole
concentration, the carriers segregate into linear walls
separating $\pi$-phase-shifted antiferromagnetic domains.
We take the appearance of one-hole line-shaped soliton for small $U/t$
in our study as an indication in favor of the former possibility.

The saddle point solution for large $U/t$
is robust and simple to understand.
Numerically we have found that
when $U/t$ is increased, the size of the
soliton is reduced.
Since the soliton has the
same quantum numbers as a {\em bare hole} in the $t$-$J$ model, we interpret it
as a finite $U$ analog of the latter, which, among other things,
also implies that the mean-field solitons obey Fermi statistics. In the
following we shall show that the low-energy fluctuations around the
saddle point solution is precisely the low-energy dynamics described by
the $t$-$J$ model:
\begin{equation}
H_{t-J} =
-t \sum_{<i,j>}
[ (1-n_{i,-\sigma})c_{i\sigma}^{\dagger} c_{j\sigma}(1-
n_{j,-\sigma}) + h.c.]
+ J \sum_{<i,j>} \vec{S}_{i} \cdot \vec{S}_{j}.
	\label{htj_2d}
\end{equation}
In the context of our above discussions the perfect N\'{e}el state
   %(say the one with electron spinning up at (0,0)?
   % WHAT DOES THIS MEAN?)
is the analog of our half-filled mean-field solution,
and $c_{i\sigma}$ acting on the latter produces our soliton. The remaining
dynamics (including the hopping of the soliton and the flipping of the
spins) described by $H_{t-J}$ are captured by fluctuations around the
saddle point.

The derivation of the effective action is similar to the 1D case.
However, unlike in 1D, the 2D solitons do have preferential
positions (i.e. the sites) even for small $U/t$. Thus
instead of a continuum description,
we use a tight-binding language to describe the soliton
hopping.
The resulting effective
action contains a O(3) non-linear sigma model
part describing the
antiferromagnetic fluctuation,
and a tight-binding spinless fermion part representing
the soliton hopping on the lattice.  Finally, unlike in 1D,
the soliton motion does couple to the
$\sigma$ model fluctuation.
This effective action is precisely the coherent-state functional-integral
action of the $t$-$J$ model in the
slave-fermion + Schwinger
boson representation.
The details are supplied in Appendix~\ref{cal_nlsm}.

\subsection{Dressing the non-topological solitons}

In the above we have argued that the effective
Hamiltonian governing the motion of the
solitons and the antiferromagnetic
fluctuations is the $t$--$J$ model.
Due to the residual coupling
between spin and charge, the soliton motion in 2D is
a highly nontrivial problem. A lot of work has been done on this
subject in this context. In particular, two of us have recently
presented
an intuitive picture that explains how the frustration of hopping in an
antiferromagnetic background renormalizes the hole bandwidth
from $\sim t$
to $\sim J$.\cite{gan}

The question of charge motion in the presence of finite doping
concentration is a much more subtle issue.
This is due to the following facts.  {\it i)\/} At finite doping
it is not clear what is the best saddle-point solution. For example
is the anti-phase domain wall solution suggested by Schultz
the correct mean-field solution for certain range of $U/t$ ?
{\it ii)\/} Even one assumes that the individual bag-like
solitons remain the lowest-energy mean-field
solution, one still has to face the fact
that in 2D, due to the residual coupling
between the solitons and
the antiferromagnetic fluctuations,
the solitons are further dressed into quasiparticles. When the
size of the one-hole quasiparticle reaches the inter-hole spacing
the meaning of a quasiparticle is no longer clear.
Moreover, the presence of finite
carrier concentration may feed back to qualitatively
change the low-energy dynamics of the
antiferromagnetic fluctuations.
Thus even though in the
case of one hole, one has
reasonable confidence that the bag-like
soliton is first formed and subsequently
is dressed into a quasiparticle,
the same conclusion should be subject
to critical scrutiny in the case of
finite doping concentration.\cite{eder94p}

With these caveats in mind let us imagine that the doping level is
relatively low so that we can view the finite doping as a collection
of one-hole quasiparticles. Let us imagine that the magnetic
long-range order is destroyed. Let us use the
antiferromagnetic fluctuations with
wavelength shorter than the magnetic
correlation length to dress the soliton\cite{gan}.
What remains is to determine the
effective theory governing the motion of
these dressed objects that includes their interaction
with the remaining long wavelength
magnetic fluctuations.

Since the residual long wavelength spin excitations
correspond to smooth fluctuations of the spin directions we
describe it by a non-linear $\sigma$
model (see Appendix~\ref{cal_nlsm} for derivation),
\begin{equation}
S_{\rm nlsm} = \frac{ \rho_{s} }{2}
\int^{\beta}_{0} d\tau
\int d^{2}r
\sum_{\mu=c_{s} \tau,x,y}
{\rm Tr} \left( \partial_{\mu} g \partial_{\mu} g^{+}
+  \sigma_{z} g \partial_{\mu} g^{+}
\sigma_{z} g \partial_{\mu} g^{+} \right)  ,
	\label{nlsm_2d}
\end{equation}
where $\rho_{s}$ is the stiffness and $c_{s}$ is the spin wave velocity.
In particular, we choose $\rho_s$ so that the resulting $\sigma$ model
is in the disordered phase. (Here we should point out the fact that how
to get a
spin liquid described by the disordered $\sigma$ model
from a microscopic spin model on a lattice is still an open issue.)
To find the effective action for
the quasiparticle propagation, we need
to generalize the
quasiparticle construction in Ref.~\onlinecite{gan}
to allow a smoothly fluctuating spin background.
Since the dressing envisioned in Ref.~\onlinecite{gan}
is a short wavelength and high-energy process,
it will not be significantly affected as long as the
spin correlation length and correlation time are
sufficiently long.
The generalized construction of the quasiparticle
for the $t$-$J$ model is described in Appendix~\ref{cal_nlsm}, where
the soliton hopping is treated exactly while the
spin-exchange part is treated on average.
The quantum spin fluctuation generates
hopping of the dressed object as a whole.
Let $i$ and $j$ be two lattice sites
with quasiparticle creation operators
$f^{\dagger}_{i}$ and $f^{\dagger}_{j}$.
The hopping matrix element between these two states is
due to the spin exchange part of the
$t$--$J$ Hamiltonian,
$H_{J}=J\sum_{<i,j>} \vec{S}_{i} \cdot \vec{S}_{j}$.
With only short range order,  the
local spin directions  at the sites
$i$ and $j$ are not necessarily
parallel to each other.
Since the quasiparticle spins are constrained to follow
the local spin directions,
the two quasiparticle states
at the sites $i$ and $j$ do not
have parallel spin directions.
As a result, the hopping amplitude acquires
a Berry phase which is expressed
in terms of the SU(2) matrix $g(i)$.
The details of the derivation are included in Appendix~\ref{cal_nlsm}.
Taking into account this additional
factor, the quasiparticle hopping
is described by,
\begin{eqnarray}
& & H_{\rm qp} =
\sum_{i \in {\cal A} , \, r=\sqrt{2}, 2}  \alpha(r)
 \left[g(i)g^{+}(i+r)\right]_{11}
f^{\dagger}_{i} f_{i+r}  +
\sum_{i \in {\cal B} , \, r=\sqrt{2}, 2}  \alpha(r)
 \left[g(i)g^{+}(i+r)\right]_{22}
f^{\dagger}_{i} f_{i+r}
	\nonumber       \\
 &  &  \hspace{.1in}
+ \;  \sum_{i \in {\cal A} ,  r=\sqrt{5},\, 3}
 \lambda(r)   \left\{
\left[g(i)g^{+}(i+r)\right]_{12}
f^{\dagger}_{i} f_{i+r}
+ h.c. \right\}  ,
		\label{hqp}
\end{eqnarray}
where the matrix subscript denotes
the respective element, and
the notation $r=\sqrt{2}$, for example,
indicates summing over all neighboring
sites of $\sqrt{2}$ units of lattice constant.
The first two terms of Eq.~(\ref{hqp})
represent the hopping among the same sublattice sites,
while the last one
describes the quasiparticle hopping term between
different sublattices.
Long range hopping
farther than three lattice spacings
 has been neglected.
We expect that the results for
the coefficients $\alpha(r)$ and $\lambda(r)$
calculated in Ref.~\onlinecite{gan}
remain a good estimate so long as the correlation length (or time) of the
antiferromagnetic order remains long.
The two parameters, $\alpha(\sqrt{2})$
and $\alpha(2)$,
describe a quasiparticle band of a
width of order $2J$,
with the minimum at $(\pi/2,\pi/2)$ and
extended van Hove regions around $(\pi,0)$ and
$(0,\pi)$.\cite{dago94,dago90,poil93,eder94,haas94,%
moreo94,bulut,germans,trug90,liu92,dago,putz}
 Alternatively, these parameters, as well as
$\lambda(r)$,
can also be taken as phenomenological parameters
of the effective theory to be determined from
fitting experimental data.

Using the representation~(\ref{g_rep})
of the SU(2) matrix, Eq.~(\ref{hqp})
can be recast as
\begin{eqnarray}
& & H_{\rm qp} =
\sum_{i \in {\cal A} , \, {2}, 2}  \alpha(r)
 \sum_{\sigma} z^{*}_{\sigma}(i) z_{\sigma}(i+r)
f^{\dagger}_{i} f_{i+r}  +
\sum_{i \in {\cal B} , \, r=\sqrt{2}, 2}  \alpha(r)
\sum_{\sigma} z_{\sigma}(i) z^{*}_{\sigma}(i+r)
f^{\dagger}_{i} f_{i+r}
	\nonumber       \\
 &  &  \hspace{.1in}
+ \;  \sum_{i \in {\cal A} ,  r=\sqrt{5},\, 3}
 \lambda(r)   \left\{
\left[
 z^{*}_{1}(i) z^{*}_{2}(j+r)
-  z^{*}_{2}(i) z^{*}_{1}(i+r)
\right]
f^{\dagger}_{i} f_{i+r}
+ h.c. \right\}  .
\end{eqnarray}
The corresponding part of the action
is
\begin{equation}
S_{\rm qp} = \int^{\beta}_{0} d\tau
\left( \sum_{i} \bar{f}_{i} \partial_{\tau}
f_{i} + H_{qp} \right) .
\end{equation}
The complete effective action is
\begin{equation}
S_{\rm eff} = S_{\rm nlsm} + S_{\rm qp} .
\end{equation}

\subsection{Closing remarks}

What can be said about the nature of the
carriers after being dressed by both short and long ranged magnetic
fluctuations? For one hole in an antiferromagnet
it is widely believed that the soliton is dressed into
a quasiparticle. Here we quote some numerical results.
For the $t$--$J$ model there exists direct diagonalization
results.\cite{dago94,dago90,poil93,eder94,haas94,moreo94}
The conclusion of these studies paints a picture in
favor of a quasiparticle with internal structure.
These results can be understood in the string picture
developed in Ref.~\onlinecite{gan}. For the 2D Hubbard
model, most numerical results are
obtained from Monte-Carlo
simulations.\cite{bulut,germans}
Unlike the direct diagonalization, these results
are for non-zero temperatures. Moreover, it is often
hard to obtain very low temperature results. In any case
the state-of-the-art low temperature
Monte-Carlo results agree with the $t$--$J$ model
diagonalization. The outcome is a
quasiparticle obeying a dispersion
relation best fitted
to\cite{haas94,bulut,germans,trug90,liu92,dago,putz}
\begin{equation}
\epsilon_{\vec{k}} = 4 \alpha(\sqrt{2})
\cos k_{x} \cos k_{y} + 2 \alpha(2)
( \cos 2k_{x} + \cos 2 k_{y}) .
\end{equation}
The two parameters $\alpha(\sqrt{2})$ and $\alpha(2)$
have also been calculated in the string picture.\cite{gan}
This result compares very favorably with
the Angle-Resolved Photoemission
results of the high temperature
superconductors near optimum
doping.\cite{shen95,campuzano,aebi94}

In the presence of a finite concentration
of holes, it is far less clear whether
the quasiparticle picture is correct.
The effective Hamiltonian, given by
Eqs.~(\ref{nlsm_2d}) and (\ref{hqp}), describes
the ultimate dressing of the quasiparticles which have
already been partially dressed by short wavelength
antiferromagnetic fluctuations. The best
chance for the relevance of the Hamiltonian
is when the magnetic correlation length (time) is long compared
with the inter-hole spacing (inverse intra-quasiparticle
excitation gap). This
Hamiltonian has been postulated
and studied by a number of
authors.\cite{weigman,shankar,wen,lee}
Here we have presented a derivation of it.

\section{Conclusion} \label{sect5}

In this paper we have demonstrated that the two
low-energy dynamical degrees of freedom in doped Hubbard
model are {\it a)} soliton charge carriers, and {\it b)} smooth
antiferromagnetic fluctuations. We have arrived
at these excitations starting from the SDW
mean-field theory. The quantum number of the
charge soliton distinguishes one dimension
from the rest. In 1D the charge solitons are
the anti-phase domain wall of the longitudinal
spin order. They carry no $\langle S^z \rangle$,
consequently
they do not suffer from further dressing by the
magnetic fluctuations. Thus in 1D the final
effective theory is a decoupled charge and spin
model. In two dimensions, the solitons
are bag-like objects and they do carry
$\langle S^z \rangle$.
Consequently, they are further dressed by
the antiferromagnetic fluctuations. The ultimate
effective theory consists of two types of
charged particles, i.e.\   with opposite
fictitious charge, coupled to fluctuating gauge
fields. The final zero temperature state
of such model is currently not known.

At last, it is important to point out that
in two space dimensions we made an
important assumption at finite
doping --- that individual solitons
are still the lowest-energy mean-field solution. If that
turns out wrong, and if the charged object has
the form of an extended domain wall,
then the discussions presented here are
irrelevant.

\acknowledgments
The authors thank  Safi Bahcall
for valuable comments
on the manuscript.

\appendix
\section{Takayama-Lin-Liu-Maki equation}  \label{app_a}

The one-dimensional Hubbard
model written in the form of
Eq.~(\ref{laglocal}) does not resemble the
Su-Schrieffer-Heeger model\cite{su79}
describing Polyacetylene
in the lattice even if
the transverse
fluctuations of the spin directions are turned off.
One difference is that the longitudinal
spin order $\Delta_{s}(i)$
does not have dynamics in Eq.~(\ref{laglocal}).
In other words, Eq.~(\ref{laglocal}) does not
contain time derivative of $\Delta_{s}(i)$.
This difference is not so important since the quantum
fluctuations will generate dynamics for $\Delta_{s}(i)$.
Another difference lies in the way how the
longitudinal spin order is coupled to the electrons.
Despite these differences,
we show in the following that the
continuum version of the saddle point equations
corresponding to Eq.~(\ref{hmfkink})
are essentially the same as the Takayama-Lin-Liu-Maki
equation\cite{tlm}
determining the
soliton solution in the continuum for
Polyacetylene.

In the continuum limit, we substitute into Eq.~(\ref{hmfkink})
\begin{equation}
c_{j} = e^{i k_{F} R_{j} } \psi_{L}(j)
+ e^{-i k_{F} R_{j} } \psi_{R}(j) ,  \hspace{.2in} R_{j}=ja,
	\label{linearize}
\end{equation}
where $\psi_{L,R}(i)$ are slowly varying fields in space,
$ \psi_{L}^{\dagger}(j)=
( \psi_{L\uparrow}^{\dagger}(j),
\psi_{L\downarrow}^{\dagger}(j))$,
$k_{F}a=\pi/2$ at the half filling,
and $a$ is the lattice constant.
The saddle point  Hamiltonian, Eq.~(\ref{hmfkink}),  becomes
\begin{eqnarray}
H_{mf} &= & -2 t a \sum_{j} \left[
\psi_{L}^{\dagger}(j) i \partial_{x} \psi_{L}(j) -
\psi_{R}^{\dagger}(j) i\partial_{x} \psi_{R}(j)  \right]
+  \sum_{j} \Delta_{c}^{(0)}(j)
\left[ \psi_{L}^{\dagger}(j) \psi_{L}(j)
+ \psi_{R}^{\dagger}(j) \psi_{R}(j) \right]
	\nonumber       \\
&  + & \sum_{j} \Delta_{s}^{(0)}(j)
\left[ \psi_{L}^{\dagger}(j)\sigma_{z} \psi_{R}(j)
+ \psi_{R}^{\dagger}(j)\sigma_{z} \psi_{L}(j) \right]
+ \frac{1}{U} \sum_{j}
\left\{ \left[ \Delta_{s}^{(0)}(j) \right]^{2}
- \left[ \Delta_{c}^{(0)}(j) \right]^{2} \right\}.
\end{eqnarray}
The discrete summation can be converted into integration
according to
\begin{equation}
 a \sum_{j} \rightarrow \int dx ;
\hspace{.2in} \frac{1}{  \sqrt{a} }
\psi_{L,R}(j) \rightarrow \psi_{L,R}(x);
\hspace{.2in} \Delta_{c,s}^{(0)}(j)
\rightarrow \Delta_{c,s}^{(0)}(x)  .
	\label{sum_to_int}
\end{equation}
The two spin components are decoupled in  the above
mean field Hamiltonian. The mean
field Hamiltonian can be rewritten as
\begin{eqnarray}
H_{mf} &=& \sum_{\alpha=\pm}
\int dx \left[ \psi^{\dagger}_{L\alpha},
\psi^{\dagger}_{R\alpha} \right]
\left[ -i v_{F} \sigma_{z} \partial_{x}
+ \Delta_{c}^{(0)}(x)
+ \alpha \Delta_{s}^{(0)}(x) \sigma_{x} \right]
\left[ \begin{array}{c}
\psi_{L\alpha}(x) \\ \psi_{R\alpha}(x) \end{array}
\right]   \     \nonumber       \\
& & + \frac{1}{Ua} \int dx \left\{
\left[ \Delta_{s}^{(0)}(x) \right]^{2}
- \left[ \Delta_{c}^{(0)}(x) \right]^{2} \right\},
\end{eqnarray}
where $v_{F}=2t a$,
$\sigma_{z}$ and $\sigma_{x}$ are Pauli matrices.
The spin index $\alpha=\pm$ corresponds to
spin up and down respectively.
This continuum version of the saddle point Hamiltonian
has the similar form as the continuum  Hamiltonian for
Polyacetylene.\cite{tlm}
Expanding the operators $\psi_{L,R}(x)$ in terms
of their eigenfunctions
\begin{eqnarray}
\psi_{L\alpha}(x) &=&
  \sum_{\kappa} u_{\kappa,\alpha}(x) \psi_{L\alpha}(\kappa) , \\
\psi_{R\alpha}(x) &=&
  \sum_{\kappa} v_{\kappa,\alpha}(x) \psi_{R\alpha}(\kappa) ,
\end{eqnarray}
the corresponding saddle point equations are
 very similar to Takayama-Lin-Liu-Maki
equation,\cite{tlm}
\begin{eqnarray}
& & -i v_{F} \partial_{x} u_{\kappa,\alpha}(x)
+ \Delta_{c}^{(0)}(x) u_{\kappa,\alpha}(x)
+  \alpha \Delta_{s}^{(0)}(x) v_{\kappa,\alpha}(x)
= \epsilon_{\kappa,\alpha} u_{\kappa,\alpha}(x)  ,  \\
& & i v_{F} \partial_{x} v_{\kappa,\alpha}(x)
+ \Delta_{c}^{(0)}(x) v_{\kappa,\alpha}(x)
+  \alpha \Delta_{s}^{(0)}(x) u_{\kappa,\alpha}(x)
= \epsilon_{\kappa,\alpha} v_{\kappa,\alpha}(x)  , \\
& & \Delta_{s}^{(0)}(x) =
- \frac{U a}{2} \sum_{\alpha=\pm}
\sum_{\kappa \in {\rm occup}}
\alpha \left[
u_{\kappa,\alpha}^{*}(x)
v_{\kappa,\alpha}(x)
+ v_{\kappa,\alpha}^{*}(x)
u_{\kappa,\alpha}(x)
\right] ,  \\
& & \Delta_{c}^{(0)}(x) =
 \frac{U a}{2}  \sum_{\alpha=\pm}
\sum_{\kappa \in {\rm occup}}
 \left[|u_{\kappa,\alpha}(x)|^{2}
+ |v_{\kappa,\alpha}(x)|^{2}
\right] .
\end{eqnarray}
The only difference is the appearance
of an additional self-consistent equation
for $\Delta_{c}^{(0)}(x)$.
This analogy
underlies the
common physics of
the holon in 1D Hubbard model and
soliton in Polyacetylene.
{}From this analogy, we identify
the counterpart of the
dimensionless electron-phonon
coupling constant $\lambda_{e-ph}$
of the Polyacetylene\cite{heeger} to be
\begin{equation}
\frac{1}{2\pi v_{F} \lambda_{e-ph} }
 = \frac{1}{Ua},
\hspace{.2in}
\Rightarrow  \hspace{.2in}
 \lambda_{e-ph}= \frac{U}{4 \pi t } .
\end{equation}
In the absence of the self-consistent equation for
$\Delta_{c}^{(0)}(x)$, the saddle point equations
have been solved in closed form giving rise  to
$\Delta_{s}^{(0)}(x)=\Delta_{s}^{(0)} \tanh(x/\xi)$.
The electron density calculated using
the eigenfunctions in the presence of the kink
has a shallow dip
at the kink position which has the form
$ 1 - 1/[ 2 \xi \cosh^{2}(x/\xi)]$.
This shallow dip causes additional scattering
effects when there is an additional equation for
$\Delta_{c}^{(0)}(x)$. Although  the
true solution in this case will be different
from that in Polyacetylene, we expect that
the difference is only
quantitative. In particular,
the change should not be significant
when $\xi \gg 1$, i.e.\  for small $U/t$.

The presence of the Hartree term distinguishes
the continuum equations in this case from
those in Polyacetylene. However, the localized
states are still degenerate in the spin index.
To see this, let us define
\begin{equation}
f_{\alpha}^{(\pm)}(x) =
u_{\alpha}(x) \pm v_{\alpha}(x) .
\end{equation}
The coupled equations for $ u_{\alpha}(x)$
and $v_{\alpha}(x)$ (the index $\kappa$ is omitted)
become
\begin{equation}
-i\left[ v_{F} \partial_{x}
\pm \alpha \Delta_{s}^{(0)}(x) \right]
f_{\alpha}^{(\pm)}(x) =
\left[ \epsilon_{\alpha} - \Delta_{c}^{(0)}(x) \right]
f_{\alpha}^{(\mp)}(x) .
\end{equation}
We see that the equations are
invariant under
$f^{(+)}_{\uparrow} \rightarrow f^{(-)}_{\downarrow}$
and $f^{(-)}_{\uparrow} \rightarrow f^{(+)}_{\downarrow}$.

\section{Calculation of the
fluctuations around the saddle point}  \label{cal_nlsm}

The general formalism of calculating Gaussian fluctuations
is as follows.
Substituting into Lagrangian~(\ref{laglocal})
the expansion
\begin{eqnarray}
\Delta_{s}(j) &=& \Delta_{s}^{(0)}(j) +
\delta \! \Delta_{s}(j)  ,  \\
\Delta_{c}(j) &=& -i \Delta_{c}^{(0)}(j) +
\delta \! \Delta_{c}(j) ,   \label{expand_c}
\end{eqnarray}
we separate the full
Lagrangian, of the form of Eq.~(\ref{laglocal})
into
a mean-field part and a fluctuation
part.
\begin{eqnarray}
{\cal L} &=& {\cal L}_{mf} + \delta{\cal L} , \\
{\cal L}_{mf} &=& \sum_{i,j} \sum_{\mu,\nu}
\bar{c}_{i,\mu}
\left( \partial_{\tau}
+ h^{(0)}_{\{i,\mu\},\{j,\nu\}}
\right) c_{j,\nu}
+ {\cal L}_{0}(\delta \! \Delta_{c},\delta \! \Delta_{s} ) ,
	\label{lmf_gen} \\
\delta{\cal L} &=& \sum_{i,j} \sum_{\mu,\nu}
\bar{c}_{i,\mu} \;
\left[ h^{(1)}( \delta \! \Delta_{c},\delta \! \Delta_{s},  g)
\right]_{\{i,\mu\},\{j,\nu\}}
\; c_{j,\nu} .
	\label{dl_gen}
\end{eqnarray}
The time-independent matrix $h^{(0)}$
can be specified once the saddle point solution is
obtained. The matrix $h^{(1)}$
depends on the fluctuating fields
$\delta \! \Delta_{c}(j)$,\  $\delta \! \Delta_{s}(j)$
  and $g$.
The quadratic part,
\begin{equation}
{\cal L}_{0}( \delta \! \Delta_{c},\delta \! \Delta_{s}  )
= (1/U)\sum_{i}\left\{ \left[
\delta \! \Delta_{s}(i) \right]^{2}
+ \left[
\delta \! \Delta_{c}(i) \right]^{2} \right\} ,
\end{equation}
is unaffected by the procedure of integrating out fermions
to obtain the effective action defined through
\begin{equation}
e^{- S_{\rm eff}( \delta \! \Delta_{c},
\delta \! \Delta_{s}  , g) }
= \int {\cal D}[c, \bar{c}] \,
e^{- \int^{\beta}_{0} d\tau {\cal L} } .
	\label{def_seff}
\end{equation}
The integration is carried out mainly
by expansion.
The first step of the
calculation is to diagonalize
${\cal L}_{mf}$.
This is achieved by a linear
transformation,
\begin{equation}
 \bar{c}_{j,\sigma}
= \sum_{\kappa} \alpha_{\kappa,j}^{(\sigma)}
\bar{\psi}_{\kappa,\sigma} .
\end{equation}
To have a Jacobian equal to 1 or to preserve the proper
commutation relation
$ \{ \psi^{\dagger}_{\kappa_{1},\sigma} ,
\psi_{\kappa_{2},\sigma} \}
=\delta_{\kappa_{1},\kappa_{2}}$ in
the operator language,
the transformation coefficients
$ \alpha_{\kappa,j}^{(\sigma)} $
must satisfy
\begin{equation}
\sum_{\kappa} \alpha_{\kappa,i}^{(\sigma)*}
\alpha_{\kappa,j}^{(\sigma)} = \delta_{i,j} .
\end{equation}
In terms of the new Grassmann variables
$\psi_{\kappa,\sigma}$ and
$\bar{\psi}_{\kappa,\sigma}$,
the mean-field Lagrangian is diagonalized,
\begin{equation}
{\cal L}_{mf} = \sum_{\kappa,\sigma}
\bar{\psi}_{\kappa,\sigma}
( \partial_{\tau} + \epsilon_{\kappa,\sigma} )
\psi_{\kappa,\sigma}
+ {\cal L}_{0}( \delta \! \Delta_{c},\delta \! \Delta_{s}  ) ,
	\label{diag_lmf}
\end{equation}
where $ \epsilon_{\kappa,\sigma} $
is the eigenvalue of the matrix $h^{(0)}$.
Then, we can represent the
interaction part of the Lagrangian
in terms of the new variables
$ \psi_{\kappa,\sigma} $ and
$ \bar{\psi}_{\kappa,\sigma} $,
\begin{equation}
\delta{\cal L} = \sum_{\kappa_{1},\kappa_{2}}
\sum_{\mu,\nu}
\bar{\psi}_{\kappa_{1},\mu}(\tau) \;
 h^{(\mu,\nu)}_{\kappa_{1},\kappa_{2}}(\tau) \;
\psi_{\kappa_{2},\nu}(\tau)  ,
\end{equation}
where
\begin{equation}
h^{(\mu,\nu)}_{\kappa_{1},\kappa_{2}}(\tau)
= \sum_{i,j}
\alpha^{(\mu)*}_{\kappa_{1},i} \;
\left[ h^{(1)}( \delta \! \Delta_{c},\delta \! \Delta_{s}, g)
\right]_{\{i,\mu\},\{j,\nu\}}
\; \alpha^{(\nu)}_{\kappa_{2},j} .
\end{equation}
In the final step, we expand the
interaction $\delta {\cal L}$ in the
partition function. To the first order,
\begin{equation}
\langle \delta{\cal L}
\rangle =
\sum_{\kappa_{1},\kappa_{2}}
\sum_{\mu,\nu}
 h^{(\mu,\nu)}_{\kappa_{1},\kappa_{2}}
\langle
\bar{\psi}_{\kappa_{1},\mu}(\tau) \;
\psi_{\kappa_{2},\nu}(\tau)
\rangle .             \label{ave_dl}
\end{equation}
In the second order of the expansion in
$\delta {\cal L}$, the generated effective action
is
\begin{equation}
S_{\rm eff}^{(2)} =
\frac{1}{2} \sum_{\nu_{n}}
\sum_{\kappa_{1},\kappa_{2}}
\sum_{\mu,\nu}
\Pi_{\mu,\nu}(\kappa_{1},
\kappa_{2},\nu_{n}) \;
h^{(\mu,\nu)}_{\kappa_{1},
\kappa_{2}}(\nu_{n}) \;
h^{(\nu,\mu)}_{\kappa_{2},
\kappa_{1}}(-\nu_{n}) ,
	\label{s2eff}
\end{equation}
where $\nu_{n}=2\pi n/\nu$ is the Bosonic
Matsubara frequency, and
$ h^{(\mu,\nu)}_{\kappa_{1},
\kappa_{2}}(\nu_{n})$
is the Fourier transform of
$ h^{(\mu,\nu)}_{\kappa_{1},
\kappa_{2}}(\tau)$.
The function
$ \Pi_{\mu,\nu}(\kappa_{1},
\kappa_{2},\nu_{n}) $
has the usual Lindhard form
\begin{equation}
\Pi_{\mu,\nu}(\kappa_{1},
\kappa_{2},\nu_{n})
= \frac{ f(\epsilon_{\kappa_{1},\mu})
- f(\epsilon_{\kappa_{2},\nu}) }
{ i\nu_{n} + \epsilon_{\kappa_{1},\mu}
- \epsilon_{\kappa_{2},\nu} } ,
\end{equation}
where $f(\epsilon)=1/(1+e^{\beta\epsilon})$ is
the Fermi-Dirac distribution function.
Since the particle-hole
excitation in the spectrum of
the mean-field state has an energy
gap, we can safely approximate
$\Pi_{\mu,\nu}(\kappa_{1},
\kappa_{2},\nu_{n})
\simeq \Pi_{\mu,\nu}(\kappa_{1},
\kappa_{2},0)$.
In the rest of this section, we shall
apply this procedure of obtaining
effective action to both 1D
and 2D Hubbard model.

\subsection{One-dimensional case}

Substituting Eqs.~(\ref{intro_Phi}) and (\ref{expand_c})
into Eq.~(\ref{laglocal}), the full 1D Lagrangian
is written as
\begin{eqnarray}
{\cal L} &=& \sum_{j} \bar{c}_{j}
 \left[ \partial_{\tau} - \widetilde{\mu }
+ i \, \delta \!\Delta_{c}(j) +
g(j) \partial_{\tau} g^{+}(j)
+ \Delta_{s}^{(0)} \cos\left(2k_{F} R_{j}
+ \Phi_{j} \right)
 \sigma_{z} \right]  c_{j}
	\nonumber               \\
& & -t \sum_{<i,j>}
\left[ \bar{c}_{i}
\, g(i)g^{+}(j) \,
c_{j} + h.c. \right]
+ \frac{ \left[\Delta_{s}^{(0)} \right]^{2} }{U}
\sum_{j} \cos^{2}(2k_{F} R_{j} + \Phi_{j} )
+ \frac{1 }{U} \sum_{j}
\left[ \delta\!\Delta_{c}(j) \right]^{2} .
	\label{laglocal1}
\end{eqnarray}
In one dimension, we can simplify
the Lagrangian by
taking the continuum limit.
The fermion fields can be linearized
around the Fermi surface, which
consists of two points in one dimension.
This is done by Eq.~(\ref{linearize})
but with $k_{F}=\pi(1-n_{h})/2$ in the general case.
Since $\Phi(i)$ and
$g(i)$ are slowly varying functions,
we can expand
\[ g(i)g^{+}(j) -1
= a g(i) \partial_{x} g^{+}(i)
+ \frac{a^{2}}{2} g(i) \partial_{x}^{2}
g^{+}(i) + \cdots,  \]
where $a$ is the lattice constant.
Using this expansion, we find
\begin{eqnarray}
& & -t \sum_{<i,j>}
\left[ \bar{c}_{i}
\, g(i)g^{+}(j) \,
c_{j} + h.c. \right]
= - v_{F} \sum_{j} \left\{
 \psi^{\dagger}_{L}(j)
\left[ i \partial_{x} + g(j) i\partial_{x} g^{+}(j)
\right] \psi_{L}(j)
- (L \rightarrow R) \right\}
	\nonumber       \\
& & \hspace{.5in}
- t a^{2} \cos(k_{F}a)
\sum_{j} \left[
 \psi^{\dagger}_{L}(j)
 g(j) \partial_{x}^{2} \left( g^{+}(j)
 \psi_{L}(j) \right)
 +(L \rightarrow R) \right],
	\label{reduce_hop}
\end{eqnarray}
where we have denoted
\begin{equation}
v_{F}=2ta \sin (k_{F}a ).
\end{equation}
This generalizes the definition of $v_{F}$
to the arbitrary electron filling factor.
Similarly, we have
\begin{equation}
\Delta_{s}^{(0)} \sum_{j} \bar{c}_{j}
\cos\left(2k_{F} R_{j}
+ \Phi_{j} \right)
 \sigma_{z}
c_{j}
=  \left\{
\begin{array}{l}
\Delta_{s}^{(0)} \cos \Phi_{j} \left[
\psi^{\dagger}_{L}(j) \sigma_{z}
\psi_{R}(j) + h.c.
\right] ,
	\; \; {\rm for} \; k_{F}a =\pi/2 ,  \\
 \frac{1}{2} \Delta_{s}^{(0)} \left[ e^{i\Phi_{j}}
\psi^{\dagger}_{L}(j) \sigma_{z} \psi_{R}(j)
+ h.c. \right] ,
\; \; {\rm for} \; k_{F}a \neq \pi/2 .
\end{array}  \right.
	\label{reduce_m0}
\end{equation}
We also note that
\begin{equation}
2 \sum_{i} \cos^{2}(2k_{F} R_{i} + \Phi_{i} )
={\cal N} +
\sum_{i} \cos( 4k_{F} R_{i} + 2 \Phi_{i})
\simeq \left\{
\begin{array}{l}
{\cal N} , \; \; {\rm for} \;
k_{F}a \neq \pi/2 ,     \\
{\cal N} + \sum_{i}
\cos (2\Phi_{i} ) ,
\; \; {\rm for} \;
k_{F}a = \pi/2 .
\end{array} \right.
	\label{vanish_m2}
\end{equation}
The other terms in Eq.~(\ref{laglocal1})
are simplified in the same way.
In the case of finite doping,
$ k_{F}a \neq \pi/2$, we can
redefine the fermion fields to
absorb the phase $ e^{i\Phi_{j}} $
accompanying $\Delta_{s}^{(0)} $
in Eq.~(\ref{reduce_m0}),
\begin{equation}
\widetilde{\psi}_{L}(j) =
 \psi_{L}(j) e^{-i \Phi_{j}/2} ,
\hspace{.2in}
\widetilde{ \psi}_{R}(j)
= \psi_{R}(j) e^{i\Phi_{j}/2} .
	\label{redef_phi}
\end{equation}
Substituting Eqs.~(\ref{reduce_hop})
through  (\ref{redef_phi})
into Eq.~(\ref{laglocal1}), we obtain for the
case of finite doping,
\begin{eqnarray}
{\cal L}_{mf}
&=& \int dx \left[ \widetilde{\psi}_{L}^{\dagger}(x),
\widetilde{\psi}_{R}^{\dagger}(x) \right]
\left[
\begin{array}{cc}
\partial_{\tau} - i v_{F} \partial_{x} &
 \frac{1}{2} \Delta_{s}^{(0)}  \sigma_{z}      \\
  \frac{1}{2} \Delta_{s}^{(0)}   \sigma_{z}      &
\partial_{\tau} + i v_{F} \partial_{x}
\end{array}
\right]
\left[
\begin{array}{l}
 \widetilde{\psi}_{L}(x) \\
\widetilde{\psi}_{R}(x)
\end{array}
\right]
	\nonumber	\\	& &
+ {\cal N} \frac{ \left[\Delta_{s}^{(0)}\right]^{2} }{2U}
+ \frac{1 }{Ua} \int dx
 \left[\delta \! \Delta_{c}(x) \right]^{2},
	\label{lmf_contin}  \\
\delta {\cal L} &=&
\int dx \left[ \widetilde{\psi}_{L}^{\dagger}(x) A_{L}(x)
\widetilde{\psi}_{L}(x) +
 \widetilde{\psi}_{R}^{\dagger}(x) A_{R}(x)
\widetilde{\psi}_{R}(x) \right]  ,
	\label{dl_contin}
\end{eqnarray}
where we have converted the
discrete summation into
integration according to Eq.~(\ref{sum_to_int})
and
\begin{equation}
g(i) \rightarrow g(x),
\hspace{.3in}
 \delta \! \Delta_{c}(i) \rightarrow
 \delta \! \Delta_{c}(x) .
\end{equation}
We have also
introduced the notations,
\begin{equation}
A_{L,R}(x) = g(x)(\partial_{\tau}
 \mp i v_{F} \partial_{x} ) g^{+}(x)
+ \frac{i}{2}( \pm \partial_{\tau}
- i v_{F} \partial_{x} )  \Phi(x)
+ i \, \delta\!\Delta_{c}(x).
\end{equation}
It is now straightforward to calculate
the Gaussian fluctuations according
to the general prescription, Eqs.~(\ref{lmf_gen})
and (\ref{dl_gen}).
We note that the site index in
Eqs.~(\ref{lmf_gen})
and (\ref{dl_gen}) corresponds
to the chirality index and the
space coordinate
in Eq.~(\ref{lmf_contin}).
The lattice summation corresponds to
\[ \sum_{i} \rightarrow
\sum_{{\rm chirality}=L,R} \int dx . \]
After Fourier transformation to the
momentum $k$ space,
Eq.~(\ref{lmf_contin}) is easily
diagonalized by a Bogoliubov transformation.
For $\delta {\cal L}$ given
by Eq.~(\ref{dl_contin}),
$\langle \delta {\cal L} \rangle =0$.
In the second order contribution from
 $\delta {\cal L}$ expansion, the
transverse spin fluctuation part
is decoupled from the rest and is given by
the non-linear sigma model, Eq.~(\ref{nlsm}).
The spin stiffness  in Eq.~(\ref{nlsm})
is given by
\begin{equation}
\rho_{s}
= \frac{2 v_{F}^{2}}{\beta}\sum_{\omega_{n}}
\int \frac{dk}{2\pi}
\left| \langle \psi^{\dagger}_{L}(k,\omega_{n})
\psi_{R}(k,\omega_{n}) \rangle
\right|^{2}
= \frac{ 2 v_{F}^{2}}{\beta}\sum_{\omega_{n}}
\int \frac{dk}{2\pi}
\left[ \frac{ \Delta_{s}^{(0)}/2 }{ \omega_{n}^{2}
+ (v_{F}k)^{2} + [\Delta_{s}^{(0)}/2]^{2} }
\right]^{2}
= \frac{v_{F}}{2\pi}.       \label{stiff_s1}
\end{equation}
Note that the Gaussian fluctuations
only give rise to the second term
of Eq.~(\ref{nlsm}). The first
term is due to the topological effect in
Eqs.~(\ref{lmf_contin}) and
(\ref{dl_contin}) which is not captured
by the expansion in
$\delta {\cal L}$.
When tracing out the fermions,
the resulting effective action
contains an imaginary term
known as the Berry phase.
In the language of
Eqs.~(\ref{lmf_contin}) and
(\ref{dl_contin}), the Berry phase
appears in the form of chiral anomaly.
Mathematically, the chiral anomaly
exists because there is
no regularization which
preserves both gauge and chiral invariance.
Physically, this Berry phase term
is the electric field experienced by
the phase of the holons $\Phi(x)$.
This Berry phase has been calculated exactly
by Nagaosa and Oshikawa\cite{nagaosa} which
is given by Eq.~(\ref{berry})
with $\theta = \pi$,
independent of doping. We shall not
repeat the calculation here.

The rest of the effective action
including up to the second order contributions
is
\begin{equation}
S_{\rm eff}(\Phi,\delta \! \Delta_{c}) =
\int_{0}^{\beta} d\tau \int dx
\left\{ \frac{ \rho_{c}^{(0)} }{2}
\left[
\frac{1}{v_{F}^{2}} (\partial_{\tau} \Phi)^{2}
+  \left( \partial_{x} \Phi
+ \frac{2i}{v_{F}} \delta \! \Delta_{c}(x) \right)^{2} \right]
+  \frac{1}{Ua}  \left[\delta\!\Delta_{c}(x)\right]^{2}
\right\},
\end{equation}
where $ \rho_{c}^{(0)} = \rho_{s}=v_{F}/2\pi$.
The $\delta \! \Delta_{c}(x) $ fluctuation is gapful and therefore
can be integrated out. The remaining low
energy mode described by $\Phi(x)$ has an
effective action
\begin{equation}
S_{\rm eff}(\Phi) =
\frac{ \rho_{c} }{2}
\int_{0}^{\beta} d\tau \int dx
 \left[
\frac{1}{v_{c}^{2}} (\partial_{\tau} \Phi)^{2}
+  \left( \partial_{x} \Phi
\right)^{2} \right] ,
	\label{s2eff_1d}
\end{equation}
where
\begin{eqnarray}
\rho_{c} &=& \rho_{c}^{(0)} \left( 1+
     \frac{Ua}{Ua + \pi v_{F} } \right)
= \frac{v_{F}}{2\pi} \left( 1+
     \frac{Ua}{Ua + \pi v_{F} } \right) ,
\label{stiff_c1}   \\
v_{c} &=& v_{F} \sqrt{ 1+
     \frac{Ua}{Ua + \pi v_{F} } } .
	\label{velo_c1}
\end{eqnarray}
We observe that
these coefficients, as well as $\rho_{s}$, are
independent of the expectation value
$  \Delta_{s}^{(0)}  $.
The increase of the charge
stiffness by renormalization indicates
that the interaction between the holons
is repulsive.
The same is true in
the doped Polyacetylene.\cite{heeger}

At half-filling, both
Eqs.~(\ref{reduce_m0}) and
(\ref{vanish_m2}) give different
results than at finite doping.
We see that a pinning potential
proportional to
$\cos 2 \Phi$ has appeared,
leading to an excitation gap
for the bosonic field $\Phi(x)$.\cite{affleck}
Thus, only spin excitations
are gapless at half-filling.
We recover the non-linear
sigma model as the effective
theory of the 1D Hubbard model
at half-filling.

Let us return to the general
case of finite doping in one dimension.
We have to further show that the various
correlation functions
can be correctly reproduced
using our effective action.
Since the effective action we have derived
has a form equivalent to the non-Abelian
bosonized action, we only need to show
that the physical fields
 entering our
effective action have the same meaning
as in the non-Abelian bosonization.
For this purpose, we can couple
external fields to the spin and
charge densities in the original
1D Hubbard model.
The rest is to show that the
generated
external field coupling terms
in our effective action have
the same forms as those in the
non-Abelian bosonization action.

Physically, we have shown in Eq.~(\ref{den_phas})
that $\partial_{x} \Phi$ has the meaning
of holon density. Since the location
of the holon coincides with
the hole, $\partial_{x} \Phi$
has the meaning of charge density.
The size of the holon is irrelevant
to the long wavelength physics.
 Formally, we can also prove that
$\partial_{x} \Phi$
has the meaning of charge density.
 We introduce an external
potential $h_{c}(x,\tau)$ and couple
it to the electron density in the original
Hubbard model,
$\sum_{i,\sigma} h_{c}(i,\tau)
c^{\dagger}_{i\sigma}
c_{i\sigma}$.
After linearization
using Eq.~(\ref{linearize}) and phase transformation
in Eq.~(\ref{redef_phi}),
it becomes
\begin{equation}
\int dx h_{c}(x,\tau)
\left[ \widetilde{\psi}_{L}^{\dagger}(x)
\widetilde{\psi}_{L}(x)
+ \widetilde{\psi}_{R}^{\dagger}(x)
\widetilde{\psi}_{R}(x) \right]
+ \int dx h_{c}(x,\tau)
\left[ \widetilde{\psi}_{R}^{\dagger}(x)
\widetilde{\psi}_{L}(x) e^{2ik_{F}x+ i\Phi(x)}
+ h.c. \right] .
	\label{c_density}
\end{equation}
Including the expression~(\ref{c_density})
in Eq.~(\ref{dl_contin}),
the first term of (\ref{c_density})
adds
the potential $h_{c}(x,\tau)$
to $\frac{i}{2}( \pm \partial_{\tau}
- i v_{F} \partial_{x} )  \Phi(x)$ in $A_{L,R}(x)$.
When we integrate out the fermionic
fields in Eq.~(\ref{lmf_contin}),
the mixing terms of type
$i h_{c}(x,\tau)( \pm \partial_{\tau}
- i v_{F} \partial_{x} )  \Phi(x) $
are generated in the second order.
This gives rise to an additional  term
$2 v_{F} h_{c}(x,\tau) \partial_{x} \Phi(x)$.
Combined with the proper prefactor,
the final term
in the effective action is indeed
$h_{c}(x,\tau)\partial_{x}\Phi(x)/\pi$.
For the second term of (\ref{c_density}),
the bilinear fermion operators
can be replaced by the saddle point
average. This leads to
$ 2 \langle  \widetilde{\psi}_{R}^{\dagger}(x)
\widetilde{\psi}_{L}(x) \rangle
\int dx h_{c}(x,\tau)
\cos[2k_{F}x+ \Phi(x)] $.
This term will give rise to the
oscillating part of the
density correlation function.

To show that the vector field
$\vec{n}$ in the effective
action represents the spin direction,
we couple a  magnetic field
$\vec{h}(x,\tau)$ to the spin operator
in the original Hubbard model
in the form of $\sum_{i} \vec{h}(i,\tau)
\cdot c^{\dagger}_{i\alpha}
\vec{\sigma}_{\alpha\beta} c_{i\beta}/2$.
Introducing this coupling term
into Eq.~(\ref{hprime}),
the last term of
Eq.~(\ref{hprime})
is changed to
$\sum_{i,\alpha\beta}
c^{\dagger}_{i\alpha}
\left[  Q_{\alpha\beta}(i)
+ h_{\alpha\beta} \right]
c_{i\beta}  + h.c.  $,
where $h_{\alpha\beta}=
(1/4)\vec{h}\cdot
\vec{\sigma}_{\alpha\beta}$
and $ Q_{\alpha\beta}(i)
= (-1)^{i} \Delta_{s}(i) \,
[ g^{+}(i) \sigma_{z} g(i) ]_{\alpha\beta} $.
We can restore the original form
for the last term of
Eq.~(\ref{hprime})
by a shift of the integration
variable: $Q_{\alpha\beta}
\rightarrow Q_{\alpha\beta}(i)
+ h_{\alpha\beta} $.
The modification is
now transferred to
the second term in
Eq.~(\ref{hprime})
which now becomes
$(1/U)\sum_{i} {\rm Tr}
|Q+h|^{2}$.
The mixing term in the expansion of
the square of $Q+h$
is the only relevant
magnetic field coupling term which
has the form
$(1/U) \sum_{i}{\rm Tr}
[\bar{Q}(i) h(i) + h.c.]
=(1/U) \sum_{i}
(-1)^{i} \Delta_{s}(i)
\vec{n}_{i} \cdot \vec{h}_{i}$
if we use the fact that
$ g^{+}(i) \sigma_{z} g(i)
=\vec{n}_{i} \cdot \vec{\sigma}$.
Upon substituting
$(-1)^{i} \Delta_{s}(i)
=\Delta_{s}^{(0)} \cos (2k_{F}R_{i}+\Phi_{i})$
into the above relation, we
obtain
$( \Delta_{s}^{(0)} /U) \sum_{i}
\cos(2k_{F} R_{i}+\Phi_{i})
 \vec{n}_{i} \cdot \vec{h}_{i}$.
The expressions for the charge and spin
density operators and their
scaling dimensions should be
compared with the non-Abelian bosonization
results.\cite{affl_1}

\subsection{Two-dimensional case}

In two dimensions,
the mean-field Lagrangian, the transformation
coefficients $\alpha^{(\sigma)}_{\kappa,j}$
and the eigenenergies
$\epsilon_{\kappa,\sigma}$ are only
obtained numerically. In practical calculations,
they are all chosen to be real numbers.
The saddle point Hamiltonian is given by
Eq.~(\ref{hmf_2d}).  The corresponding matrix
elements are
\begin{equation}
h^{(0)}_{\{i,\mu\},\{j,\nu\}}
= -t \, \delta_{\mu,\nu} \; \sum_{r =\pm x, \pm y}
\delta_{i-j,r} +
\delta_{\mu,\nu} \,
\delta_{i,j} \left[
\Delta_{c}^{(0)}(i) +
{\rm sgn} \mu \, (-1)^{i_{x}+i_{y}}
\Delta_{s}^{(0)}(i)
\right],
\end{equation}
where $i,j$ label the lattice sites, and $\mu,\nu$ are
the spin indices.
The typical $\Delta_{s}^{(0)}(i)$
and $\Delta_{c}^{(0)}(i)$ for the $14 \times 14$
lattice are shown in Fig.~\ref{bag_profile}
which is the solution of Eqs.~(\ref{consis1})
and (\ref{sdeq_c_2d}).
The transformation coefficients
$\alpha^{(\sigma)}_{\kappa,j}$
are just the eigenvectors
obtained by numerically diagonalizing
$h^{(0)}_{\{i,\mu\},\{j,\nu\}}$.
The matrix element in the interaction part
of the Lagrangian, Eq.~(\ref{dl_gen}), is
identified to be
\begin{eqnarray}
\left[ h^{(1)}( \delta \! \Delta_{c}(j),
 \delta \! \Delta_{s}(j), g)
\right]_{\{l,\mu\},\{j,\nu\}}
&=& \delta_{l,j}  \,
\left[ g(j) \partial_{\tau}
g^{+}(j) \right]_{\mu,\nu}
-t \, \sum_{r=\pm x, \pm y}
\delta_{l-j,r}  \,
\left[ g(l)g^{+}(j)-1 \right]_{\mu,\nu}
	\nonumber       \\
& & \, + \delta_{l,j} \, \delta_{\mu,\nu}
 \left[ i \, \delta \! \Delta_{c}(j) + {\rm sgn} \mu \,
 (-1)^{j_{x}+j_{y}} \,  \delta \! \Delta_{s}(j) \right],
\end{eqnarray}
where $\delta \! \Delta_{c}(j)$,
 $\delta \! \Delta_{s}(j)$ and the SU(2) matrix $g(j)$
are the fluctuating fields.

The first order term in the expansion of
the effective action, Eq.~(\ref{ave_dl}),
does not contribute anything interesting
since we know that the Berry phase
term is canceled out in 2D.\cite{berry}
In the second order,  Eq.~(\ref{s2eff}),
we obtain an action in the variables
$\delta \! \Delta_{c}(j)$,\
 $\delta \! \Delta_{s}(j)$ and  $g(j)$.
Unlike in the 1D calculation,
the coefficients in the effective
action have spatial variation
since we are perturbing around
a saddle point solution which is
not uniform in space.
The translational invariance will
be restored when the soliton motion
is included.
Thus, we can simply perform  the
spatial average of
the coefficients in $S^{(2)}_{\rm eff}$.
Because the soliton has
a finite spatial size, although small for
$U/t \gg 1$, the effective
action $S^{(2)}_{\rm eff}$
contains nonlocal terms,
such as
$ \sum_{i,i'} f(i-i')
{\rm Tr} \left[ g(i)\partial_{\tau}g^{+}(i)
g(i')\partial_{\tau}g^{+}(i') \right] $,
where the function $f(r)$ decreases exponentially
with the distance $r$.
In the momentum space, this nonlocality
corresponds to a momentum dependence
of the coupling constant which is the
spatial Fourier
transform of $f(i-i')$.
If only the long wavelength
behavior
is concerned,
this nonlocality can be neglected
by approximating
$ \sum_{i,i'} f(i-i')
{\rm Tr} \left[ g(i)\partial_{\tau}g^{+}(i)
g(i')\partial_{\tau}g^{+}(i') \right]
\simeq \sum_{i} \left[ \, \sum_{i'} f(i-i') \right]
{\rm Tr} \left[ g(i)\partial_{\tau}g^{+}(i)
g(i)\partial_{\tau}g^{+}(i) \right] $.
This is equivalent to neglecting
the momentum dependence of the coupling
constant and taking its value at
the momentum $\vec{k}=0$.
After the two steps described above,
i.e.\  averaging out the spatial
nonuniformity and neglecting the nonlocality
of the coefficients in  $S^{(2)}_{\rm eff}$,
the $g(j)$ fluctuations
are decoupled from $\delta \! \Delta_{c}(j)$ and
 $\delta \! \Delta_{s}(j)$.
 Both $\delta \! \Delta_{c}(j)$ and
 $\delta \! \Delta_{s}(j)$
fluctuations are gapful. So they can
be neglected.
The $g(j)$ fluctuations are described
by the non-linear sigma model,
Eq.~(\ref{nlsm_2d}),
with the stiffness constant given by
\begin{equation}
\rho_{s} =
\frac{t^{2}}{2{\cal N}}
\sum_{\kappa,\kappa'}
\frac{ \theta(\epsilon_{\kappa,\uparrow})
- \theta(\epsilon_{\kappa',\downarrow}) }
{\epsilon_{\kappa,\uparrow} -
\epsilon_{\kappa',\downarrow}  }
\left[ \sum_{i}
\alpha^{(\uparrow)}_{\kappa,i}
\left( \alpha^{(\downarrow)}_{\kappa',i+x}
+ \alpha^{(\downarrow)}_{\kappa',i-x} \right)
\right]
\left[ \sum_{i}
\alpha^{(\downarrow)}_{\kappa',i}
\left( \alpha^{(\uparrow)}_{\kappa,i+x}
+ \alpha^{(\uparrow)}_{\kappa,i-x} \right)
	\label{stiff_s2d}
\right]   ,
\end{equation}
where ${\cal N}$ is the number of
lattice sites and $\theta(\epsilon)$
is the step function.
The spin wave velocity $c_{s}$
is given by
\begin{equation}
\frac{1}{c_{s}^{2}} =
\frac{1}{2{\cal N} \rho_{s} }
\sum_{\kappa,\kappa'}
\frac{ \theta(\epsilon_{\kappa,\uparrow})
- \theta(\epsilon_{\kappa',\downarrow}) }
{\epsilon_{\kappa,\uparrow} -
\epsilon_{\kappa',\downarrow}  }
\left( \sum_{i}
\alpha^{(\uparrow)}_{\kappa,i}
\alpha^{(\downarrow)}_{\kappa',i} \right)^{2} .
	\label{speed_s2d}
\end{equation}
These expressions can be evaluated
using the numerical soliton solution
in $14 \times 14$ lattice for the one-hole problem.
The difference of the stiffness from that
at half-filling can be viewed as the
renormalization due to the hole.
In  the presence of a finite density
of holes, the naive generalization is
to multiply the one-hole
renormalization
by the hole density.
However, caution must be exercised
if one attempts to apply Eqs.~(\ref{stiff_s2d})
and Eqs.~(\ref{speed_s2d}) to
large $U/t$ ratios.
This is because these two expressions
involve not only the ground state but also
all electron-hole pair excited states.
It is reasonable to expect that the mean field
solution is a good approximation to
the ground state since the self-consistency
has been achieved. But the excited states
constructed by creating electron-hole pairs
in the mean-field spectrum may not
be a good approximation of the true
charge-excited states of the full Hubbard model.

So far we have calculated the Gaussian
 fluctuations around the saddle point
solution,  i.e.\    the soliton.
Therefore, only small fluctuations
have been
taken into account.
Since there are many degenerate saddle
point solutions, corresponding to the
different locations of the solitons,
we must sum up the fluctuation
contributions around
all saddle point solutions.
Furthermore, we must also include
the tunneling processes between
the different saddle point solutions.
In terms of the eigenoperators
of the saddle point Hamiltonian
corresponding to (\ref{diag_lmf})
whose $\Delta_{s}^{(0)}(i)$ profile has a dip
located at the site $i_{0}$,
the soliton creation operator
$ f_{i_{0}}^{\dagger} $ is defined
through
\begin{equation}
|{\rm Bag}(i_{0}) \rangle
= \prod_{\rm Lower \; bands}
\psi_{\kappa,\sigma}^{\dagger}
|{\rm Vacuum} \, \rangle
\stackrel{\rm define}{=}
f_{i_{0}}^{\dagger} |{\rm SDW} \rangle ,
	\label{construc_bag}
\end{equation}
where $|{\rm SDW} \rangle$ is the state without
hole.  The soliton operator has finite
overlap with the bare hole operator
in Eq.~(\ref{lmf_gen}),
\begin{equation}
c_{i_{0}, \sigma_{0}}
=\sqrt{Z_{\rm bag}} \;  f^{\dagger}_{i_{0}} +
\cdots .
\end{equation}
For a given site, $\sigma_{0}$
can only take one value
depending on the direction of the
magnetization on the site $i_{0}$.
Note that the SU(2) rotational symmetry
is restored by the fluctuating $g(i)$.
Numerically, it is
straightforward to verify, in
the $14 \times 14$ lattice,
 that
$\sum_{i,\sigma} \delta
\langle c^{\dagger}_{i\sigma}
c_{i\sigma} \rangle =-1$
and $ | \sum_{i} \delta
\langle S^{z}_{i} \rangle | = 1/2$.
Since all the spatial variations
take place locally around the center of
the soliton, we have to assign
these quantum numbers to
the soliton.
Similarly, we can take another pair of
$\Delta_{s}^{(0)}(i)$ and $\Delta_{c}^{(0)}(i)$
 profiles with exactly the same
spatial variation except that the dip
is centered on a nearest neighbor
site of $i_{0}$.
Thus, we construct another soliton
in complete analogy with
Eq.~(\ref{construc_bag})
and define the soliton creation
operator
$f_{j_{0}}^{\dagger}$ on
the nearest neighbor site
of $i_{0}$.
If we repeat the calculations
determining the quantum numbers of
the soliton, we find
$\sum_{i,\sigma} \delta
\langle c^{\dagger}_{i\sigma}
c_{i\sigma} \rangle =-1$, same as before,
but $  \sum_{i} \delta
\langle S^{z}_{i} \rangle $
has changed sign.
So the two solitons constructed
above have opposite $S^{z}$
quantum numbers.
The interaction part
of the Hamiltonian corresponding to Eq.~(\ref{dl_gen})
has a nonzero matrix element
between these two solitons:
\begin{equation}
\langle {\rm Bag}(j_{0}) |
\delta {\cal L}
|{\rm Bag}(i_{0}) \rangle
= - \widetilde{t} \,
\left[ g(i_{0}) g^{+}(j_{0}) \right]_{12} ,
\end{equation}
where $\widetilde{t} =t
\times |{\rm overlap \; integral}|$.
This corresponds
to a hopping term of the soliton
in the effective action~(\ref{def_seff}),
\begin{equation}
 - \widetilde{t} \,
f_{i_{0}} f^{\dagger}_{j_{0}}
\left[ g(i_{0}) g^{+}(j_{0}) \right]_{12} .
	\label{coh_hop}
\end{equation}
Using the representation
of the SU(2) matrix $g(i)$, i.e.\
Eq.~(\ref{g_rep}), we can rewrite
$\left[ g(i_{0}) g^{+}(j_{0}) \right]_{12}
= z^{*}_{1}(i_{0}) z^{*}_{2}(j_{0})
-  z^{*}_{2}(i_{0}) z^{*}_{1}(j_{0}) $.
Thus, we see that the tunneling term
between neighboring solitons
is precisely the hopping term
of the $t$--$J$ model when it is
written in terms of the slave fermions
and Schwinger bosons in conjugate
representations for the two
sublattices. Apart from renormalizing $t$
to $\widetilde{t}$, the only difference is that
the correlation between the hole location and
the spin magnitude has been included
in the mean-field calculation of the Hubbard model.
This correlation has been
taken into account in determining the size
of the soliton.

So far,
we have eliminated the high
energy processes of order
$U$.  The new
``elementary particle''
is the soliton.
The effective action at this level
consists of Eq.~(\ref{nlsm_2d})
and (\ref{coh_hop}) which are equivalent
to the $t$-$J$ model~(\ref{htj_2d}).
The remaining energy
scales are $t$ (we redenote $\widetilde{t}$ by $t$) and $J$.
In the next step, we want
to construct the quasiparticle from the soliton
and eliminate the energy scale
$t$. Two of the
present authors have recently
shown how to
construct a quasiparticle state
in the  N\'{e}el state
and calculated the quasiparticle
dispersion and spectral weight.\cite{gan}
When the N\'{e}el order is destroyed
by finite doping,
 we shall construct the quasiparticle
state from a smoothly varying
antiferromagnetic
background. The spirit is the
adiabatic approximation:
The hopping of the solitons
is a local and high-energy process
while the low-energy spin excitations
are described by slow variation of
the spin directions.
Without holes, the
short wavelength
spin fluctuations
cost a lot of energy
and can be eliminated. The result
is the renormalization of the
coefficients in the non-linear
sigma model.
In the doped case,
the energy cost of
short wavelength
spin fluctuations near the solitons
is compensated by the energy gain resulting
from the hopping of the solitons.
Therefore,
short wavelength
spin fluctuations  cannot
simply be eliminated as in the
undoped case.

Let $| SRO \rangle$ be
a typical spin configuration
in a short range
antiferromagnetically correlated state
in an arbitrary region of
space, this configuration can be
obtained from the perfectly ordered
state by applying smoothly varying
SU(2) rotations:
\begin{equation}
| SRO \rangle = \prod_{i} g(i) |SDW \rangle .
\end{equation}
We apply the soliton creation
operator to this configuration
to generate a soliton
at an arbitrary site
\begin{equation}
| i_{0}, l=0 \rangle =
f^{\dagger}_{i_{0}} | SRO \rangle  ,
\end{equation}
where we used the notation
$l=0$ to indicate that
the number of highly frustrated
spins is zero,  i.e.\
there is no spin which is almost
parallel to its neighboring spins.
Applying on $| i_{0}, l=0 \rangle$
the effective Hamiltonian for
the solitons,  i.e.\
the $t$--$J$ Hamiltonian~(\ref{htj_2d}),
a new state $| i_{0}, l=1 \rangle$ is generated,
\begin{equation}
H_{t-J} | i_{0}, l=0 \rangle
= -2 t \, | i_{0}, l=1 \rangle +
\langle i_{0}, l=0 | H_{t-J}
| i_{0}, l=0 \rangle  \;
| i_{0}, l=0 \rangle  + \cdots,
	\label{string0}
\end{equation}
where  $| i_{0}, l=1 \rangle$ is
generated by the hopping term of
the $t$--$J$ model. It is a superposition
of four configurations in which
the soliton  has made one hop
from the original site $i_{0}$
to anyone of the four neighbors.
The notation $l=1$ indicates
that there is one
spin, at the site $i_{0}$,
which is frustrated
with its neighbors.
This is the generalization of the
string state of one unit length
discussed in Ref.~\onlinecite{gan} to
the smoothly varying antiferromagnetic
background.
The other generalized string states,
$ | i_{0}, l \rangle $ for $l \geq 1$,
are constructed by repeatedly
applying
the $t$--$J$ Hamiltonian to $| i_{0},l=1 \rangle$.
\begin{eqnarray}
& & H_{t-J} | i_{0}, l=1 \rangle = - 2 t
| i_{0}, l=0 \rangle
- \sqrt{3} \, t   | i_{0}, l=2 \rangle
	\nonumber	\\    & & \hspace{1in}
+ \langle i_{0}, l=1 | H_{t-J}
| i_{0}, l=1 \rangle  \;
| i_{0}, l=1 \rangle  + \cdots,
	\label{string1}  \\
& & H_{t-J} | i_{0}, l \rangle
= - \sqrt{3} \, t  \left(
| i_{0}, l-1 \rangle
+  | i_{0}, l+1 \rangle \right)
+ \langle i_{0}, l | H_{t-J}
| i_{0}, l \rangle  \;
| i_{0}, l \rangle  + \cdots,
\hspace{.2in} l \geq 2 .
	\label{stringl}
\end{eqnarray}
Eqs.~(\ref{string0}), (\ref{string1})
and (\ref{stringl}) are the effective
representation of the $t$--$J$ model
in the determination of the quasiparticle
structure. Diagonalizing these
equations with appropriate
$ \langle i_{0}, l | H_{t-J}
| i_{0}, l \rangle  $
gives rise to a quasiparticle
at the site $i_{0}$
of the form,
\begin{equation}
| i_{0} \rangle =
\sum_{l=0}^{\infty}  u_{l} \,
| i_{0}, l \rangle   ,    \label{qp_state}
\end{equation}
where $u_{l}$ are numerical
coefficients which decrease exponentially
as $l$ increases.
The quasiparticles
located at other lattice sites
are constructed in the same way.

The formation of the quasiparticle further
renormalizes the coefficients in the non-linear
sigma model. Although it is not easy to
calculate the renormalization accurately, it
is rather easy to see that hole (or soliton)
hopping inside the quasiparticle reduces the
spin stiffness. The spin stiffness is roughly
proportional to the longitudinal spin magnitude.
The soliton hopping
inside the quasiparticle reduces the average
spin magnitude $\langle S \rangle$
by an amount of
$(2 \langle l \rangle  \langle S \rangle +1)$
times the density of holes,
where $\langle l \rangle $ is the average
string length inside the quasiparticle, Eq.~(\ref{qp_state}).
For $J/t < 1$, we have $\langle l \rangle > 1$.

Now, we show that
the hopping amplitude of the
quasiparticle constructed in the way
described above
contains a Berry phase factor.
Let $ |i \rangle$ and  $ |j \rangle$
be two quasiparticle states at two
different sites of the same sublattice,
$i-j=x+y$ or $2x$,  we want to show that
the quasiparticle hopping amplitude
in the tight-binding approximation
contains the appropriate phase factor, i.e.,
\begin{equation}
\langle i| H_{t-J}  | j \rangle
- \langle i| H_{t-J}  | i \rangle
\langle i|  j \rangle
=  \alpha(i-j)
\langle \vec{n}_{i} | \vec{n}_{j} \rangle ,
\end{equation}
where $\alpha(r)$ is a real coefficient, and
$| \vec{n}_{j} \rangle $ is the coherent
spin state defined through
$ \vec{n} \cdot \vec{S} | \vec{n} \rangle
= S |\vec{n} \rangle$.
Since the quasiparticle hopping is generated
by the quantum spin fluctuations, it suffices
to show that the matrix element
of the Heisenberg operator
$ \langle i| H_{J}  | j \rangle $
contains the appropriate
phase factor.
Using Eq.~(\ref{qp_state}),
we write
\begin{equation}
\langle i| H_{J}  | j \rangle
= \sum_{l,l'} u_{l} u_{l'}
\langle i, l| H_{J}  | j, l' \rangle  .
\end{equation}
The state $| j, l' \rangle $ is a superposition
of spin configurations; Each of them contains a
string of length $l'$.
So the matrix element
$\langle i, l| H_{J}  | j, l' \rangle $
is a sum of  matrix elements
of $ H_{J}$ between two string configurations.
We need to show that the matrix element
of $ H_{J}$ between any two string configurations
contain the phase factor
$\langle \vec{n}_{i} | \vec{n}_{j} \rangle$.
In the Bethe lattice
approximation(without winding path), the
spin exchange operator
in $H_{J}$ generates quasiparticle hopping
by chopping off the strings in
$| j, l' \rangle$ by two units.\cite{gan}
In all these processes, we can explicitly verify
that $\langle i, l| H_{J}  | j, l' \rangle
\propto  \langle \vec{n}_{i} | \vec{n}_{j} \rangle $.
A general example is shown in Fig.~\ref{bethe_lat}.
The existence of winding paths (Trugman process) leads
to additional contributions to the matrix elements
$\langle i, l| H_{J}  | j, l' \rangle$.
In general, these contributions are not
exactly  proportional
to $\langle \vec{n}_{i} | \vec{n}_{j} \rangle $.
An example is shown in Fig.~\ref{trugman}.
Although not exact, the phase factor is
approximately proportional to
$ \langle \vec{n}_{i} | \vec{n}_{j} \rangle $,
 for smoothly varying spin directions.
Furthermore, the Trugman process
occurs only for $l+l' \geq 6$ and therefore it
is multiplied by a small
coefficient $u_{l} u_{l'}$.
Thus, we have shown that
the quasiparticle hopping
amplitude acquires the Berry phase factor
if the spin directions vary smoothly
in space and time.
The Berry phase factor can be expressed in
terms of the SU(2) matrices $g(i)$.
Depending on which sublattice $i$ and $j$ belong
to, the Berry phase factor
$ \langle \vec{n}_{i} | \vec{n}_{j} \rangle $
is either
$\left[ g(i) g^{+}(j)  \right]_{11} $
or $\left[ g(i) g^{+}(j)  \right]_{22} $.
For the quasiparticle hopping between different
sublattices, $i-j=2x+y$ or $3x$,
the appropriate phase factor
of the hopping amplitude can be
determined in the same way.

       %  \end{multicols}

\clearpage

\pagebreak

\begin{figure}
%\centerline{\epsfxsize=13cm\epsfbox{fig1.ps} }

%\bigskip
\caption{
Cartoon picture of the quasiparticles,
represented by the small spheres, surfing
on the fluctuating
membrane whose normal directions
define the local directions
of the magnetic moments.
The north pole of the spheres
represents
the quasiparticle spin direction
which must be perpendicular
to the membrane. The normal
vectors of the membrane
form a sphere of unit radius.
The quasiparticle traveling
a closed path picks up
a phase that is
proportional to the area
in the unit sphere enclosed
by the normal vectors
of the membrane
along the quasiparticle trace.
}
\label{surf}
\end{figure}

%\pagebreak

\begin{figure}
%\centerline{\epsfxsize=13cm\epsfbox{fig2.ps} }

%\bigskip
\caption{
Hartree-Fock solution of a holon in a 100-site
chain with open boundary condition for $U/t=8/3$.
The zigzag of the spin density near the two ends of the chain
is due to the open boundary condition and has no
influence on the holon solution at the middle of the chain.
}
 \label{var_kink_e}
\end{figure}

%\pagebreak

\begin{figure}
%\centerline{\epsfysize=12cm\epsfbox{fig3.ps}  }

%\bigskip
\caption{
The one-electron spectrum  of the
Hamiltonian~\protect{(\ref{hmfkink})}
for $U/t=8/3$.
For clarity the whole spectrum has shifted up
by amount $0.148$,
i.e.\ taking $\widetilde{\mu}=\mu-U/2=-0.148$
in Hamiltonian~\protect{(\ref{hmfkink})}.
There are two localized states
inside  the Mott-Hubbard gap.
Each localized state
contains $50\%$ spectral weight transferred
from each of the two Hubbard bands.
}
	\label{kink_spec}
\end{figure}

%\pagebreak

\begin{figure}
%\centerline{\epsfxsize=13cm\epsfbox{fig4.ps} }

%\bigskip
\caption{
The wave functions of the midgap localized states
of the one-electron spectrum
around the holon for $U/t=8/3$.
The lines are the guide to the eye.
}
\label{loc_wavefunc}
\end{figure}

%\pagebreak

\begin{figure}
%\centerline{\epsfxsize=13cm\epsfbox{fig5.ps}  }

%\bigskip
\caption{
Fitting the spin and charge densities of
a holon to analytic functions for $U/t=8/3$.
The points are the numerical solution.
The lines are
$ \langle n_{i}\rangle=1-
1/(2\xi_{c}\cosh^{2}[(i-i_{0})/\xi_{c}] )$
with $\xi_{c}=3.209$
for the charge density,
and $2(-1)^{i}\langle S_{i}^{z} \rangle
= 0.545 \tanh[(i-i_{0})/\xi_{s}]$ with
$\xi_{s}=2.978$ for the spin density.
}
\label{verif_consis}
\end{figure}

%\pagebreak

\begin{figure}
%\centerline{\epsfxsize=13cm\epsfbox{fig6.ps}  }

%\bigskip
\caption{
The magnetization
profile around the kink of optimal size for
$U/t=8/3$.
The length of the arrows is proportional to the
spin expectation value $\langle S^{z}_{i} \rangle$.
}
\label{kink_profile}
\end{figure}

%\pagebreak

\begin{figure}
%\centerline{\epsfxsize=13cm\epsfbox{fig7.ps}  }

%\bigskip
\caption{
The coarse grained
magnetization
profile of a holon.
The length of the arrows is proportional to the
spin expectation value
$\langle (S^{z}_{i}+ S^{z}_{i+1})/2 \rangle$,
and the arrows are
placed at the middle of the lattice links.
After coarse graining, the magnetizations
are nonzero only around the kink
and they sum to zero. Thus, the holon
carries $S^{z}=0$.
}
\label{ave_profile}
\end{figure}

%\pagebreak

\begin{figure}
%\centerline{\epsfxsize=13cm\epsfbox{fig8.ps} }

%\bigskip
\caption{
Variational energy of the
kink when an additional electron
is filled into the midgap states for $U/t=8/3$.
The kink as a local deformation becomes unstable
and decays into a smooth twist.
The reference energy is
that of the SDW state at half-filling of the
100-site chain with open boundary condition.
The lines are the guide to the eye.
}
\label{var_xi_lamb}
\end{figure}

%\pagebreak

\begin{figure}
%\centerline{\epsfxsize=13cm\epsfbox{fig9.ps} }

%\bigskip
\caption{
Sketch of a spinon excitation
which is a smooth twist of the
local magnetizations.
}
\label{spinon_profile}
\end{figure}

%\pagebreak

\begin{figure}
%\centerline{\epsfxsize=8cm\epsfbox{fig10a.ps} }

%\centerline{\epsfxsize=8cm\epsfbox{fig10b.ps} }

%\bigskip
\caption{
The profile of
the spin and charge densities
$2 (-1)^{i_{x}+i_{y}} \langle S^{z}_{i} \rangle$
and  $1-\langle n_{i} \rangle$
of the spin bag solution
for $U/t=6$ in a $14 \times 14$ lattice with
periodical boundary condition.
Only the region near the bag is shown.
The lattice spacing is the same as the mesh unit.
}
\label{bag_profile}
\end{figure}

%\pagebreak

\begin{figure}
%\centerline{\epsfysize=12cm\epsfbox{fig11.ps}  }

%\bigskip
\caption{
The one-electron spectrum
of the spin bag solution
for $U/t=6$.
There are two localized states
inside the Mott-Hubbard gap.
Each midgap level can be thought
of as being pushed out of
one of the two Hubbard bands.
}
	\label{bag_levels}
\end{figure}

%\pagebreak

\begin{figure}
%\centerline{\epsfxsize=13cm\epsfbox{fig12.ps} }

%\bigskip
\caption{
The variations of the local magnetizations
from the half-filled case for $U/t=6$.  The length
of arrows is proportional to
$\delta \langle S^{z} \rangle$.
The length of the center arrow has
be reduced by a factor of $10$ for clarity.
The changes
are localized around the center
of the spin bag, and they sum to $\pm 1/2$.
The sign
depends on which sublattice the center
of the spin bag resides.
Thus, the spin bag carries
$S^{z}= \pm 1/2$, in
contrast to the holon in 1D
which carries $S^{z}=0$.
Note that the variation of the staggered magnetization
has the same meaning as the coarse grained
ones shown in Fig.~\protect{\ref{ave_profile}}
since the coarse graining at half-filling
over each plaquette
yields zero everywhere.
}
\label{bag_change}
\end{figure}

%\begin{figure}
%\caption{
%The spin profile of the line shaped
%spin bag solution
%for $U/t=4/3$ in a $14 \times 14$ lattice with
%periodical boundary condition.
%The values are
%$2 (-1)^{i_{x}+i_{y}} \langle S^{z}_{i} \rangle$.
%}
%\label{strip}
%\end{figure}

%\pagebreak

\begin{figure}
%\centerline{\epsfxsize=13cm\epsfbox{fig13.ps} }

%\bigskip
\caption{
A general example of
the processes generating quasiparticle
hopping
with no winding path.
Starting from
configuration $A$, we remove a hole
either at the site $i$ or $j$,
marked by the circles, to generate $B$ or $B'$.
The hole in $B$ hops three times
to generate $C$
which contains
a string of length 3
marked by the dashed line.
$C'$ is generated from $B'$
by one hop of the hole.
$C$ and $C'$ belong to
$|i, l=3\rangle$ and
$|j,l'=1\rangle$ respectively.
$H_{J}$
contains the spin exchange operator
that can exchange the first
two spins along the string in $C$.
The resulting configuration after
exchange is shown in
$D$ together with $C'$. The matrix
element
$\langle i,l=3| H_{J} |j,l'=1\rangle$
is proportional to
$\langle \vec{n}_{i} | \vec{n}_{j} \rangle$.
}
\label{bethe_lat}
\end{figure}

%\pagebreak

\begin{figure}
%\centerline{\epsfxsize=13cm\epsfbox{fig14.ps} }

%\bigskip
\caption{
An example of
the Trugman process generating quasiparticle
hopping.
The configurations $B$ and  $B'$ are generated
by removing a hole in $A$
either at the site $a$ or $e$,
marked by the circles.
 $C$ is a configuration in
$|a, l=6\rangle$.
$H_{J}$
can exchange spins $e$ and $b$
in configuration $C$, generating $D$.
The contribution of this example to
$\langle a,l=6| H_{J} |e,l'=0\rangle$
is given by the overlap
between $D$ an $B'$, and is proportional to
$\langle \vec{n}_{a} | \vec{n}_{c} \rangle
\langle \vec{n}_{c} | \vec{n}_{e} \rangle
\simeq
\langle \vec{n}_{a} | \vec{n}_{e} \rangle $
in the smooth antiferromagnetic background.
}
\label{trugman}
\end{figure}


\begin{references}
\bibitem{note}
At half-filling, the Hubbard gap in one dimension
should be proportional to $\exp(-2\pi t/U)$
for $U/t \ll 1$. This is equivalent to
have the correct weak-coupling scaling equation.
The factorization of the Hubbard interaction
in terms of the vector spin field in
Reference~\protect{\onlinecite{nagaosa}} results in
a Hubbard gap that is exponentially wrong,
as noted by Schulz\protect{\cite{schulz90}}.
\end{references}
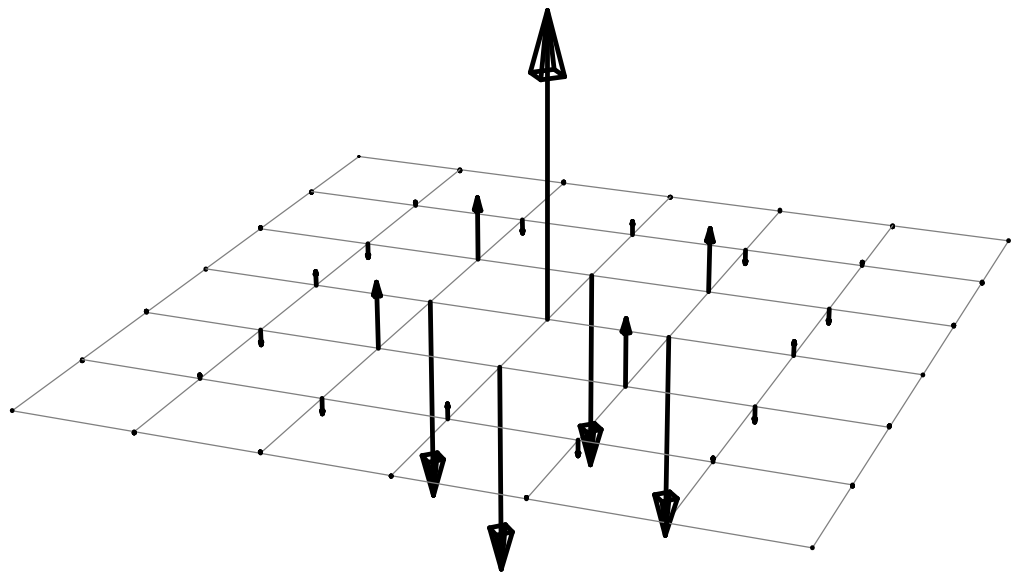
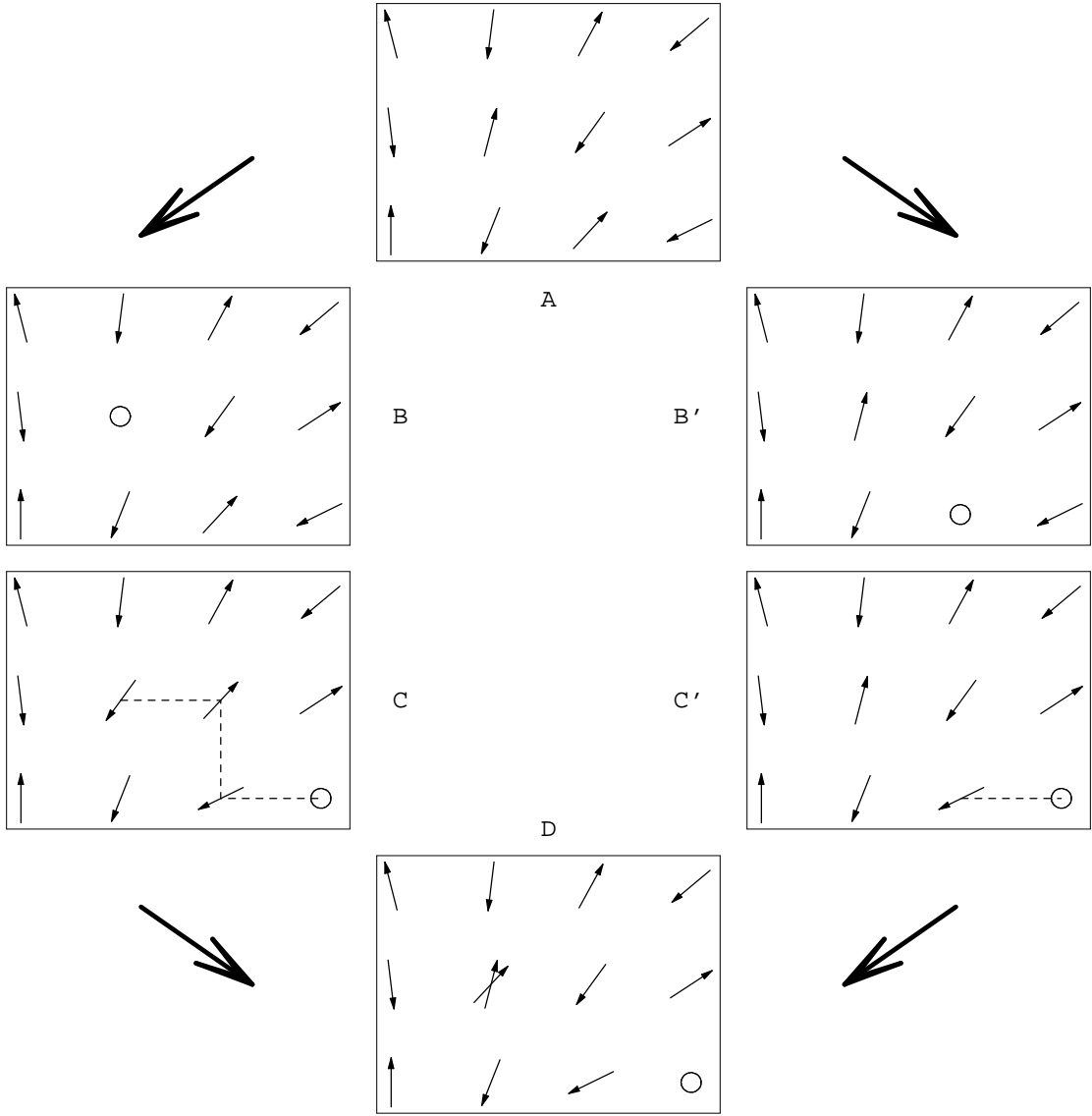
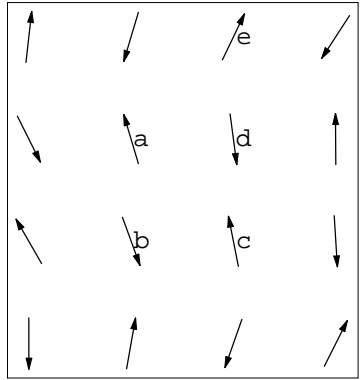
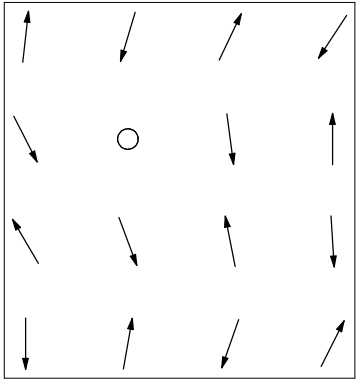
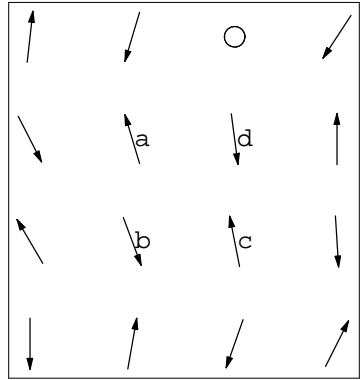
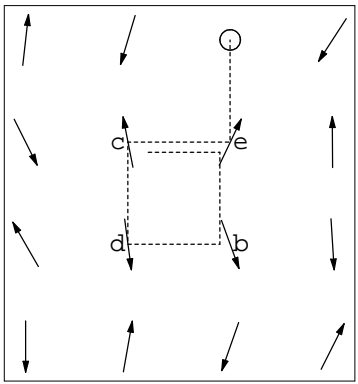
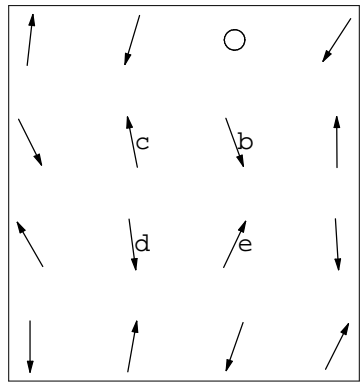
\end{document}